\def\mearth{{\rm\,M_{Earth}}}
\def\rearth{{\rm\,R_{Earth}}}
\def\msun{{\rm\,M_{Sun}}}

\def\gsim{~\rlap{$>$}{\lower 1.0ex\hbox{$\sim$}}}
\def\lsim{~\rlap{$<$}{\lower 1.0ex\hbox{$\sim$}}}

\def\eg{{\it e.g.\ }}

\def\ie{{\it i.e.\ }}

\def\kepler{{\it Kepler}}

\documentclass{emulateapj}
\usepackage{graphicx}
\usepackage{amsmath}
\begin{document}

\title{A Method to Identify the Boundary Between Rocky and Gaseous Exoplanets from Tidal Theory and Transit Durations}
\author{Rory Barnes\altaffilmark{1,2,3}}
\altaffiltext{1}{Astronomy Department, University of Washington, Box 951580, Seattle, WA 98195}
\altaffiltext{2}{NASA Astrobiology Institute -- Virtual Planetary
Laboratory Lead Team, USA}
\altaffiltext{3}{E-mail: rory@astro.washington.edu}

\begin{abstract}
The determination of an exoplanet as rocky is critical for the
assessment of planetary habitability. Observationally, the number of
small-radius, transiting planets with accompanying mass measurements
is insufficient for a robust determination of the transitional mass or
radius. Theoretically, models predict that rocky planets can grow
large enough to become gas giants when they reach $\sim 10~\mearth$,
but the transitional mass remains unknown. Here I show how transit
data, interpreted in the context of tidal theory, can reveal the
critical radius that separates rocky and gaseous exoplanets. Standard
tidal models predict that rocky exoplanets' orbits are tidally
circularized much more rapidly than gaseous bodies', suggesting the
former will tend to be found on circular orbits at larger semi-major
axes than the latter. Well-sampled transits can provide a minimum
eccentricity of the orbit, allowing a measurement of this differential
circularization. I show that this effect should be present in the data
from the \kepler~spacecraft, but is not apparent. Instead, it appears
that there is no evidence of tidal circularization at any planetary
radius, probably because the publicly-available data, particularly the
impact parameters, are not accurate enough. I also review the bias in
the transit duration toward values that are smaller than that of
planets on circular orbits, stressing that the azimuthal velocity of
the planet determines the transit duration. The ensemble of
\kepler~planet candidates may be able to determine the critical radius
between rocky and gaseous exoplanets, tidal dissipation as a function
of planetary radius, and discriminate between tidal models.
\end{abstract}

\section{Introduction}

Planetary habitability is a complex function of orbits, composition,
atmospheric evolution and geophysical processes. Most searches for
habitable environments begin with the comparison of a planet's orbit
relative to the host star's habitable zone (HZ), the region around a
star for which an Earth-like planet can support water on its surface
\citep{Dole64,Kasting93,Selsis07,Kopparapu13}. A critical feature of
this definition is the presence of a solid surface, \ie the planet
must be rocky. Hence, this determination is also crucial in the
identification of potentially habitable environments.

Unfortunately a robust and universal definition of the boundary
between rocky and gaseous worlds has remained elusive. Initially,
research was theoretical and identified the mass of a solid body, a
``core,'' that was large enough to permit the capture of
protoplanetary gas
\cite[\eg][]{Pollack96,Ikoma01,Hubickyj05,Guillot05,Militzer08,Lissauer09,Movshovitz10}. These results
showed that a wide range of ``critical masses'' are possible, from
$\lsim 1 - 20~\mearth$. Undoubtedly, the actual critical mass is a
function of the local protoplanetary disk's properties
(e.g. temperature and viscosity), and one should expect the critical
mass to vary from system to system. However, there should exist a
maximum mass (or radius) below which planets are terrestrial-like
because small mass planets do not possess enough gravitational force
to hold on to hydrogen and helium. In this study, I examine the
possibility that this boundary can be identified with transit data
coupled with expectations from tidal theory.

Transiting exoplanets offer the opportunity to measure both the
planetary radius and mass. The planetary radius can be constrained if the
stellar radius is known, usually determined from spectral information
\citep{Torres10,Everett13}, but sometimes directly through interferometric
observations \citep{vonBraun11,Boyajian12}. The mass can then be
measured through radial velocity measurements, which no longer suffer
from the mass-inclination degeneracy as the viewing geometry is known
\citep[\eg][]{Batalha11}. Recently, \cite{Mislis12} demonstrated that \kepler~light curves can reveal the mass of a close-in and massive planets via ellipsoidal variations. In multiple planet systems, masses can also be
measured by transit timing variations \citep{Agol05,HolmanMurray05},
as for Kepler-9 \citep{Holman10} and Kepler-11
\citep{Lissauer11}. However, for many transiting planets, these
methods may not be available, as terrestrial planets tend to produce small
radial velocity and timing variation signals. Thus, direct
measurements of the masses of small planets orbiting FGK stars (0.7 --
$1.4~\msun$) will be challenging.

The transit detection method is biased toward the discovery of planets on
close-in orbits, $\lsim 0.1$~AU for FGK stars. NASA's
\kepler~spacecraft has detected over 1000 planet candidates in this
range, opening up the possibility that statistical analyses of these
(uninhabitable) planets could reveal the critical radius between
terrestrial and gaseous planets, $R_{crit}$. As theoretical models of
planet formation have found that the critical mass and radius lies
near $10~\mearth$ and $2~\rearth$, planets that are smaller could be
rocky. Although few data points exist, it does appear that $R_{crit} <
2~\rearth$, \eg Kepler-10 b with mass $M_p = 4.56~\mearth$ and radius
$R_p = 1.42~\rearth$ \citep{Batalha11} and CoRoT-7 b with $M_p <
8~\mearth$ and $R_p = 1.58~\rearth$
\citep{Leger09,Queloz09,FerrazMello11}. Here I will call rocky planets
larger than the Earth ``super-Earths'', and gaseous planets less than
$10~\mearth$ ``mini-Neptunes.''

Planets amenable to both transit and radial velocity measurements tend
to lie close to their stars, which is a different environment than any
of the planets in our Solar System. These planets are subjected to
more radiation, stellar outbursts \citep{Ribas05}, and tidal effects
\citep{Rasio96,Jackson08_evol}. While these effects can act in tandem
\citep{Jackson10}, here I only consider the tidal effects, as in
isolation they can be used to calculate $R_{crit}$.

The key feature is that the expected rates of tidal dissipation in
terrestrial planets is orders of magnitude larger than for gaseous
worlds. In the classical equilibrium tidal theory
\citep{Darwin1880,GoldreichSoter66,Hut81,Jackson08_evol}, tidal dissipation is
inversely proportional to the ``tidal quality factor'' $Q$. For rocky
planets in our Solar System, $Q_r \sim 100$ \citep{Yoder95,Henning09},
whereas giants have $Q_g = 10^4 - 10^7$
\citep{GoldreichSoter66,Yoder95,ZhangHamilton08,Lainey12}, with a
traditional value of $10^6$
\citep{Rasio96,Jackson08_evol,Jackson09}. For a typical star-planet
configuration, tides usually damp both orbital eccentricity and
semi-major axis, hence the different dissipation rates result in
different damping timescales. In other words, terrestrial planets
should circularize more quickly and/or at larger separations than
gaseous planets. This discrepancy could reveal the value of
$R_{crit}$, and also $Q_r$ and $Q_g$, all of which are still poorly
constrained observationally.

In the next section, I describe the tidal models used in this study
and the expected orbits of exoplanets in the \kepler~field of view. In
$\S$~3 I describe how transit data can constrain eccentricity through
the transit duration and basic orbital mechanics, including an
extended discussion of observational biases. In $\S$~4 I describe how
I produce hypothetical distributions of transiting planets that
undergo tidal evolution. In $\S$~5 I present the results and find that
the different orbital evolutions of rocky and gaseous exoplanets
should be detectable, but are not apparent in publicly-available
\kepler~data. I find that those data contain a
distribution of impact parameters that are inconsistent with isotropic
orbits, suggesting a systematic error may be present. In $\S$~6 I
discuss the results and finally in $\S$~7 I draw my conclusions.

\section{Tidal Theory}

\begin{figure}[h]
\centering
\begin{minipage}{3in}
\includegraphics[angle=90,width=3in]{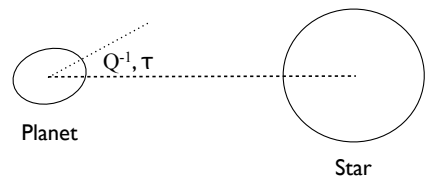}
\end{minipage}
\caption{\label{fig:tides}Schematic of the position of a planetary tidal bulge due to a star. Dissipative processes inside the planet prevent the bulge, marked by the dotted line, from pointing toward the center of the perturber, shown by the dashed line. Instead the bulge points away by an angle that is either inversely proportional to the tidal $Q$ (for the CPL model), or proportional to the time lag $\tau$ (for the CTL model).}
\end{figure}

For my calculations of tidal evolution, I employ ``equilibrium
tide'' models, originally conceived by
\cite{Darwin1880}. This model assumes the gravitational
potential of a perturber can be expressed as the sum of Legendre
polynomials (\ie surface waves) and that the elongated equilibrium
shape of the perturbed body is slightly misaligned with respect to the
line that connects the two centers of mass, see
Fig.~\ref{fig:tides}. This misalignment, which only occurs for non-zero eccentricity and obliquity, is due to dissipative
processes within the deformed body and leads to a secular evolution of
the orbit as well as the spin angular momenta of the two bodies. These
assumptions produce 6 coupled, non-linear differential equations, but
note that the model is, in fact, linear in the sense that there is no
coupling between the surface waves which sum to the equilibrium
shape. Considerable research has explored the validity and subtleties
of the equilibrium tide model
\cite[\eg][]{Hut81,FerrazMello08,Wisdom08,EfroimskyWilliams09,Leconte10}. For this
investigation, I will use the models and nomenclature of
\cite{Heller11} and \cite{Barnes13}, which are summarized below.

\subsection{The Constant--Phase--Lag Model}

In the ``constant-phase-lag'' (CPL) model of tidal evolution, the
angle between the line connecting the centers of mass and the tidal
bulge is constant. Thus, the planet responds to the perturber like a
damped, driven harmonic oscillator. The CPL model is commonly used in
planetary studies
\citep[e.g.][]{GoldreichSoter66,Greenberg09}. Under this assumption, and ignoring the effect of obliquity, the
evolution is described by the following equations

\begin{equation}\label{eq:e_cpl}
  \frac{\mathrm{d}e}{\mathrm{d}t} \ = \ - \frac{ae}{8 G M_1 M_2}
  \sum\limits_{i = 1}^2Z'_i \Bigg(2\varepsilon_{0,i} - \frac{49}{2}\varepsilon_{1,i} + \frac{1}{2}\varepsilon_{2,i} + 3\varepsilon_{5,i}\Bigg)
\end{equation}

\noindent and

\begin{equation}\label{eq:a_cpl}
  \frac{\mathrm{d}a}{\mathrm{d}t} \ = \ \frac{a^2}{4 G M_1 M_2}
  \sum\limits_{i = 1}^2 Z'_i  \ {\Bigg(} 4\varepsilon_{0,i} + e^2{\Big [} -20\varepsilon_{0,i} + \frac{147}{2}\varepsilon_{1,i} + \nonumber \frac{1}{2}\varepsilon_{2,i} - 3\varepsilon_{5,i} {\Big ]}{\Bigg )} 
\end{equation}

\noindent where $e$ is eccentricity, $t$ is time, $a$ is semi-major axis, $G$ is Newton's gravitational constant, $M_1$ and $M_2$ are the two masses,
and $R_1$ and $R_2$ are the two radii. The
quantity $Z'_i$ is

\begin{equation}\label{eq:Zp}
Z'_i \equiv 3 G^2 k_{2,i} M_j^2 (M_i+M_j) \frac{R_i^5}{a^9} \ \frac{1}{n Q_i} \,
\end{equation}

\noindent where $k_{2,i}$ are the Love numbers of order 2, $n$ is the mean motion, and $Q_i$ are the tidal quality factors. The signs of the phase lags
are

\begin{equation}\label{eq:epsilon}
\begin{array}{l}
\varepsilon_{0,i} = \Sigma(2 \omega_i - 2 n)\\
\varepsilon_{1,i} = \Sigma(2 \omega_i - 3 n)\\
\varepsilon_{2,i} = \Sigma(2 \omega_i - n)\\
\varepsilon_{5,i} = \Sigma(n)\\
\varepsilon_{8,i} = \Sigma(\omega_i - 2 n)\\
\varepsilon_{9,i} = \Sigma(\omega_i) \ ,\\
\end{array}
\end{equation}

\noindent where $\omega_i$ is the rotational frequency of the $i$th body, which I force to be the equilibrium frequency, $(1+9.5e^2)n$. $\Sigma(x)$ is the sign of any physical quantity $x$, and thus
$\Sigma(x)~=~+1, -1$, or 0.

\subsection{The Constant--Time--Lag Model}

The constant-time-lag (CTL) model assumes that the time interval
between the passage of the perturber and the tidal bulge is a constant
value, $\tau$. This assumption allows the tidal response to be
continuous over a wide range of frequencies, unlike the CPL
model. But, if the phase lag is a function of the forcing frequency,
then the system is no longer analogous to a damped driven harmonic
oscillator. Therefore, this model should only be used over a narrow
range of frequencies, see \cite{Greenberg09}. Ignoring obliquity, the
orbital evolution is described by the following equations:

\begin{equation} \label{eq:e_ctl}
  \frac{\mathrm{d}e}{\mathrm{d}t} \ = \ \frac{11 ae}{2 G M_1 M_2}
  \sum\limits_{i = 1}^2Z_i \Bigg(\frac{f_4(e)}{\beta^{10}(e)}  \frac{\omega_i}{n} -\frac{18}{11} \frac{f_3(e)}{\beta^{13}(e)}\Bigg)
\end{equation}

\begin{equation}\label{eq:a_ctl}
  \frac{\mathrm{d}a}{\mathrm{d}t} \ = \  \frac{2 a^2}{G M_1 M_2}
  \sum\limits_{i = 1}^2 Z_i \Bigg(\frac{f_2(e)}{\beta^{12}(e)} \frac{\omega_i}{n} - \frac{f_1(e)}{\beta^{15}(e)}\Bigg)
\end{equation}

\noindent where

\begin{equation}\label{eq:Z}
 Z_i \equiv 3 G^2 k_{2,i} M_j^2 (M_i+M_j) \frac{R_i^5}{a^9} \ \tau_i \ ,
\end{equation}

\noindent and 

\begin{equation}\label{eq:f_e}
\begin{array}{l}
\beta(e) = \sqrt{1-e^2},\\
f_1(e) = 1 + \frac{31}{2} e^2 + \frac{255}{8} e^4 + \frac{185}{16} e^6 + \frac{2
5}{
64} e^8,\\
f_2(e) = 1 + \frac{15}{2} e^2 + \frac{45}{8} e^4 \ \ + \frac{5}{16} e^6,\\
f_3(e) = 1 + \frac{15}{4} e^2 + \frac{15}{8} e^4 \ \ + \frac{5}{64} e^6,\\
f_4(e) = 1 + \frac{3}{2} e^2 \ \ + \frac{1}{8} e^4,\\
f_5(e) = 1 + 3 e^2 \ \ \ + \frac{3}{8} e^4.
\end{array}
\end{equation}
As in the CPL model, I force the rotational frequency to equal the equilibrium frequency, which is $(1+6e^2)n$ in the CTL model.

There is no general conversion between $ Q_\mathrm{p}$ and
$\tau_\mathrm{p}$. Only if $e~=~0$ (and the obliquity is 0 or $\pi$), when
merely a single tidal lag angle exists, then
\begin{equation}\label{eq:qtau}
Q_\mathrm{p}~\approx~1/(2|n-\omega_\mathrm{p}|\tau_\mathrm{p}),
\end{equation}
as long as $n-\omega_\mathrm{p}$ remains unchanged. Hence, the
canonical values of the dissipation parameters for dry, rocky planets
in the Solar System, $Q~=~100$ \citep{GoldreichSoter66} and $\tau =
64$~s
\citep{Lambeck77,Barnes13}, are not necessarily equivalent. Hence, the results
for the tidal evolution will intrinsically differ among the CPL and
the CTL model, even though both choices are common for the respective
model. While tidal dissipation in super-Earths remains observationally
unconstrained, here I will assume that it is similar to the rocky
bodies in the Solar System.

\subsection{Differential Circularization of Super-Earths and Mini-Neptunes}

\begin{figure}[h]
\centering
\begin{minipage}{2.7in}
\resizebox{2.7in}{!}{\includegraphics{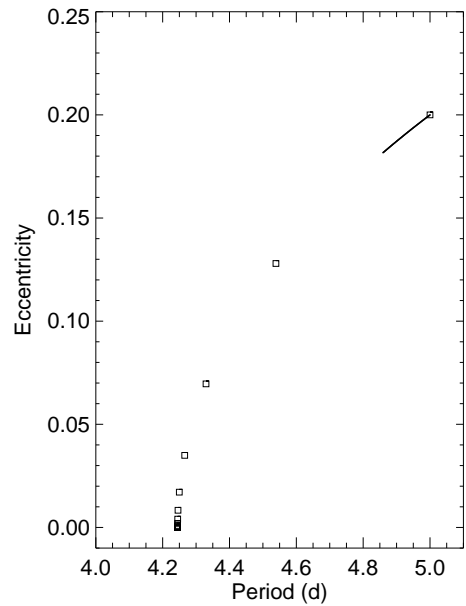}} 
\end{minipage}
\begin{minipage}{2.7in}
\caption{\label{fig:compareQ}Comparison of the tidal evolution of a gaseous $2~\rearth$
mini-Neptune (solid line; for 10 Gyr) and a rocky $2~\rearth$ super-Earth
(open squares; in 100 Myr intervals). Both planets begin with an
orbital period of 5 days and and eccentricity of 0.2 and tidally evolve down and to the left. The mini-Neptune
experiences little orbital evolution, but the super-Earth circularizes
in about 1 Gyr. This discrepancy is due to the 4 orders of magnitude
difference in tidal dissipation between gaseous and rocky planets.}
\end{minipage}
\end{figure}

As described above, most previous research predicts several orders of
magnitude difference in $Q$ or $\tau$ between terrestrial and giant
planets. As an example in Fig.~\ref{fig:compareQ}, I simulate the
evolution of two hypothetical equal-radius planets with different compositions that formed with
identical orbits (5 day period, $e = 0.2$) around identical stars. The line shows 10 Gyr of CPL
tidal evolution of a $2~\rearth$ planet with a density of 1 g/cm$^3$
(\ie a gaseous $3.8~\mearth$ mini-Neptune) and a tidal $Q$ of
$10^6$~\citep[see e.g.][]{GoldreichSoter66,Jackson08_evol}, while the
filled circles are the orbit of a $2~\rearth$ planet with a mass of
$10~\mearth$ and a tidal $Q$ of 100 (\ie a super-Earth) every 100
Myr. The super-Earth circularizes in about 1 Gyr; the mini-Neptune
barely evolves, even over 10 Gyr. This discrepancy is despite that the
equilibrium tidal models predict evolution scales with planetary mass
-- the large difference between the $Q$s dominates.

\section{The Transit Duration Anomaly}

In this section I review and revise previous work on the ``transit
duration anomaly'' (TDA), defined here as the ratio of the observed
transit duration to the duration if the orbit were circular. This
parameter has gone by several names in the literature, such as the
``transit duration deviation'' \citep{Kane12}, and the
``photoeccentric effect'' \citep{DawsonJohnson12}. Here I use the name
proposed by \cite{Plavchan12}\footnote{Note that
\cite{Plavchan12} define the TDA as the ratio of the observed duration
to that of a planet on an eccentric orbit that transits the stellar
equator.}, as in celestial mechanics the term ``anomaly'' refers to a
parameter's value relative to pericenter, \eg the ``true anomaly'' is
the difference between a planet's true longitude and its longitude of
pericenter. The analogy is not perfect, as the TDA is not measured
relative to the duration at pericenter, but rather to the duration due
to a circular orbit. Nonetheless, ``anomaly'' captures the fact that
the duration is set by the longitude relative to pericenter, the true
anomaly $\theta$, as shown below.

In this section, I first review how the TDA can be used to determine
the minimum eccentricity of an orbit. Then I review the biases
implicit in TDA measurements and update previous results.

\subsection{Determination of the Minimum Eccentricity}

\begin{figure}[h]
\centering
\begin{minipage}{3in}
\includegraphics[angle=90,width=3in]{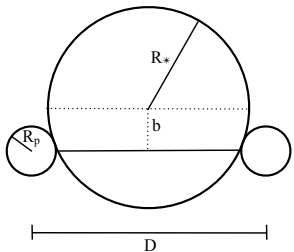}
\end{minipage}
\caption{\label{fig:transit}Schematic of a planetary transit. The planet has a radius $R_p$, and the star $R_*$. The planet crosses the stellar disk a projected distance $b$ from disk center. The distance the planet travels during transit is $D$, the distance between the center of the planet at first and fourth contacts.}
\end{figure}

Transit data coupled with knowledge of semi-major axis enable the
imposition of a lower bound on the eccentricity $e_{min}$ (Jason
W. Barnes 2007; Ford et al.~2008). The determination of $e_{min}$
requires knowledge of both the physical size of the orbit, as well as
a precise determination of the orbital period $P$, planetary and
stellar radii ($R_p$ and $R_*$), and the impact parameter $b$, see
Fig.~\ref{fig:transit}. If the transit is well-sampled, then these
parameters can be obtained \citep{MandelAgol02}.

The transit duration is the time required for a planet to traverse the
disk of its parent star, and to first order is:
\begin{equation}\label{eq:duration}
T = \frac {D}{v_{sky}} = \frac{2(\sqrt{(R_*+R_p)^2 - b^2}}{v_{sky}},
\end{equation}
where $v_{sky}$ is the azimuthal velocity, \ie the instantaneous velocity of the planet in the
plane of the sky. Although several different definitions of the
duration are possible \citep{Kipping10}, I choose this definition to match the \kepler~public data. On a circular orbit, the
azimuthal velocity is constant and equal to the orbital
velocity. Therefore the duration for a circular orbit is
\begin{equation}\label{eq:durcirc}
T_c = \frac{\sqrt{(R_*+R_p)^2 - b^2}}{\pi a}P,
\end{equation}
where $P$ is the orbital period.  For an eccentric orbit the orbital
velocity is a function of longitude (Kepler's 2nd Law), and is given
by
\begin{equation}\label{eq:velocity}
v = v_c\sqrt{\frac{1 + 2e\cos\theta + e^2}{1-e^2}},
\end{equation}
where $e$ is the eccentricity and $v_c$ is the circular
velocity. Finally, from classical mechanics, the azimuthal velocity is
\begin{equation}\label{eq:vtheta}
v_{sky} = v_c\frac{1+e\cos\theta}{\sqrt{1-e^2}}. 
\end{equation}
From transit data alone, the value of $\theta$ is unknown, and hence so is $e$.

However, one can exploit the difference between $T$ and $T_c$ to
obtain a minimum value of the eccentricity, $e_{min}$ \citep{Barnes07}. The situation is somewhat complicated because $T$ can be larger
or smaller than $T_c$ depending on $\theta$. If the planet is close to
apoapse, $T > T_c$, while near periapse $T < T_c$. To derive
$e_{min}$, one must assume that $\theta = 0$ or $\pi$. While the
velocity could be larger at some other position in the orbit, the
maximum deviation from the circular velocity is at least as large as
the measured velocity, and hence $e$ must be at least a certain
value. If I define the TDA as
\begin{equation}\label{eq:delta}
\Delta \equiv T/T_c = v_c/v_{sky} = \frac{\sqrt{1-e^2}}{1+e\cos\theta},
\end{equation}
then
\begin{equation}\label{eq:emin}
e_{min} = \left|\frac{\Delta^2 - 1}{\Delta^2+1}\right|
\end{equation}
is the minimum eccentricity permitted by transit data.

Several studies invoked the orbital velocity instead of the sky
velocity
\cite[\eg][]{TingleySackett05,Burke08,Kipping10}. However, J.~W.~Barnes (2007) correctly noted that
$T$ is actually a function of the azimuthal velocity, which
equals the orbital velocity at pericenter and apocenter. For many
other studies, the assumed form of the velocity is unclear. As
transits are a photometric phenomenon, unaffected by the component of
the velocity along the line of sight, the projection of the orbital
velocity into the sky plane, $v_{sky}$, is the appropriate
choice. This implies that previous studies are only approximately
correct. As I show in the next section, using $v_c$ will only amount
to a small error.

The TDA has been used in several studies to constrain the eccentricity
distribution, often with the assumption that the impact parameter is
unknown, as proposed by \cite{Ford08}. In that case, one can only use
those systems in which $T > T_c$ for a central transit ($b=0$) to
estimate $e_{min}$. \cite{Moorhead11} analyzed the first 3 quarters of
\kepler~data and found that the KOIs appeared to be consistent with a
mean eccentricity near 0.2. They also found that eccentricities appear
to be large regardless of orbital period, and that small planets tend
to have larger eccentricities. More recent
work has failed to determine if the \kepler~eccentricity distribution
is consistent with the radial velocity planets
\citep{Plavchan12,Kane12}. These studies were limited by the number of
known candidates, as well as the relatively poor characterization of
the transits themselves. Note that \cite{Kane12} used the difference
between $T$ and $T_c$, rather than the quotient, to model
eccentricity. \cite{DawsonJohnson12} demonstrated that in some cases
careful statistical analyses of transit data can provide constraints
on the actual eccentricity, especially if $\Delta$ deviates
significantly from unity.

\subsection{Observational Biases in the Transit Duration Anomaly}

\begin{figure*} 
\begin{tabular}{cc}
\includegraphics[width=0.49\textwidth]{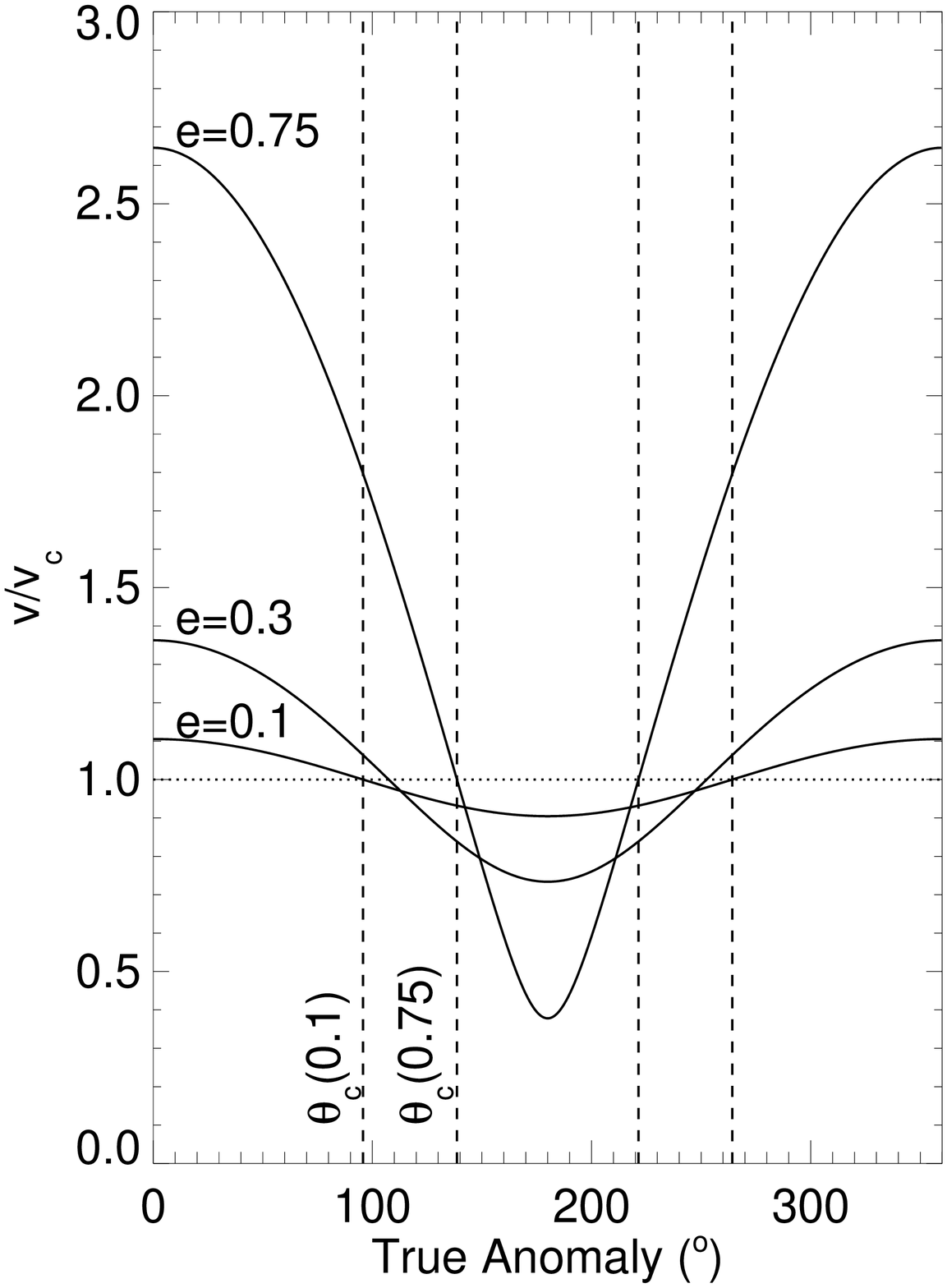} &
\includegraphics[width=0.49\textwidth]{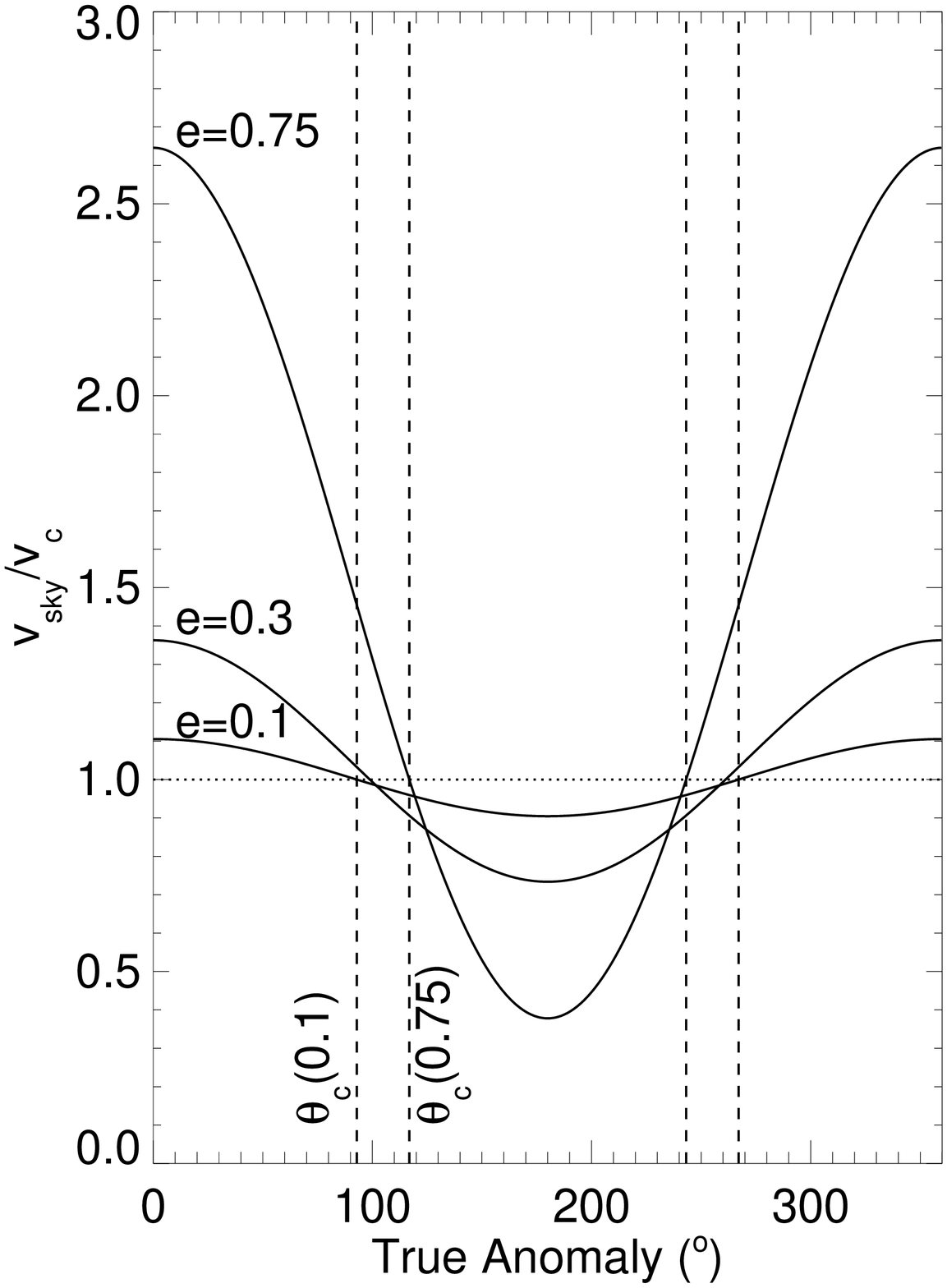}
\end{tabular}
\caption{Velocity relative to the circular velocity as a function of true anomaly for three different eccentricities. The dotted line shows the circular velocity. Dashed lines show the values of true anomaly at which the velocity equals the circular velocity. \textit{Left:} Orbital velocity. \textit{Right:} Velocity in the plane of the sky.} 
\label{fig:vtheta}
\end{figure*}

To further elucidate the TDA, as well as to revisit previous results that
may have invoked an inappropriate definition, in this section I review
the geometry and biases associated with it. In Fig.~\ref{fig:vtheta} I
show the orbital and and azimuthal velocities as a function of true
anomaly for three different eccentricities. The differences between
the panels are subtle at low $e$, but can be significant for large
$e$.

\begin{figure*} 
\begin{tabular}{cc}
\includegraphics[width=0.49\textwidth]{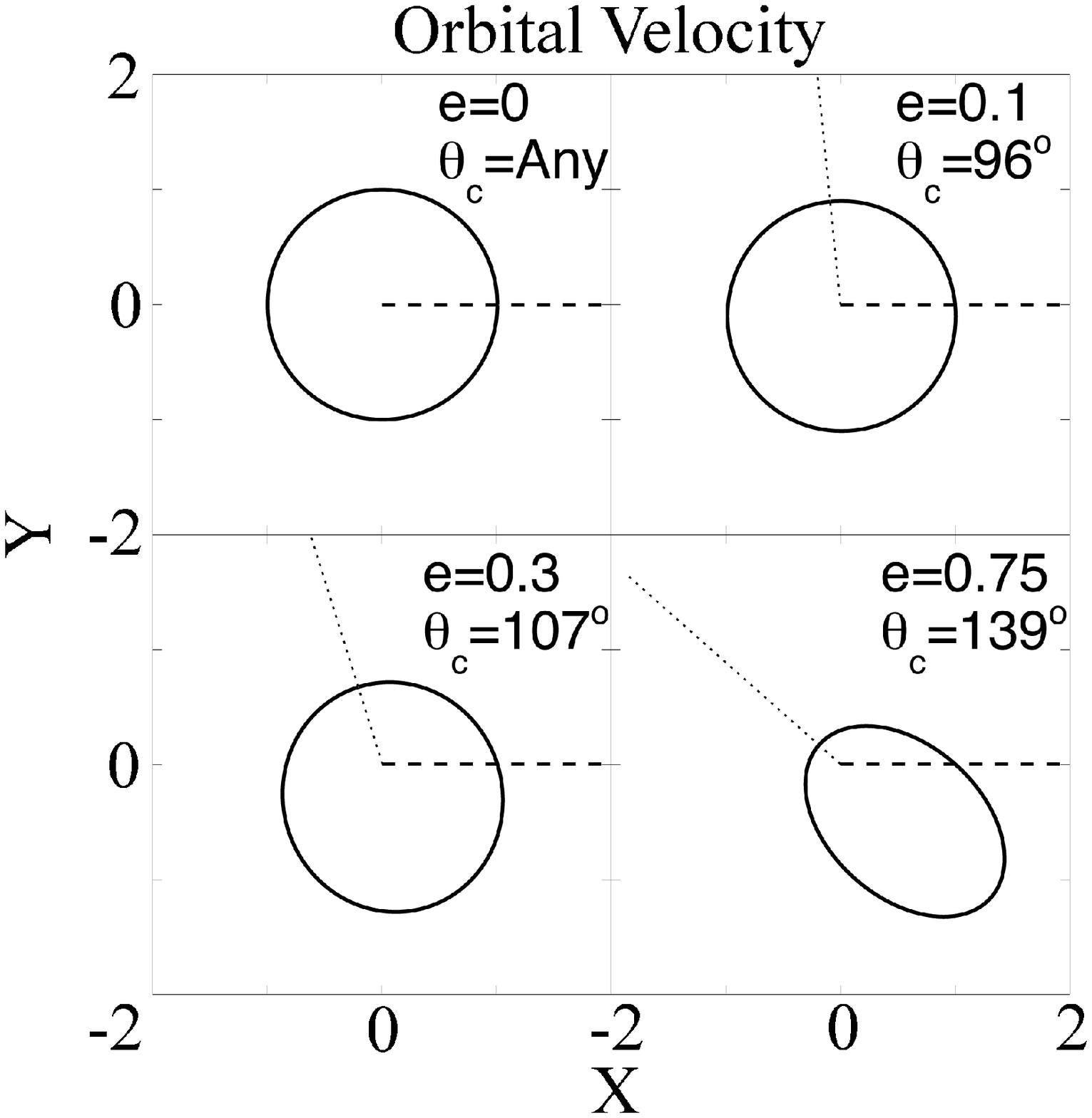} &
\includegraphics[width=0.49\textwidth]{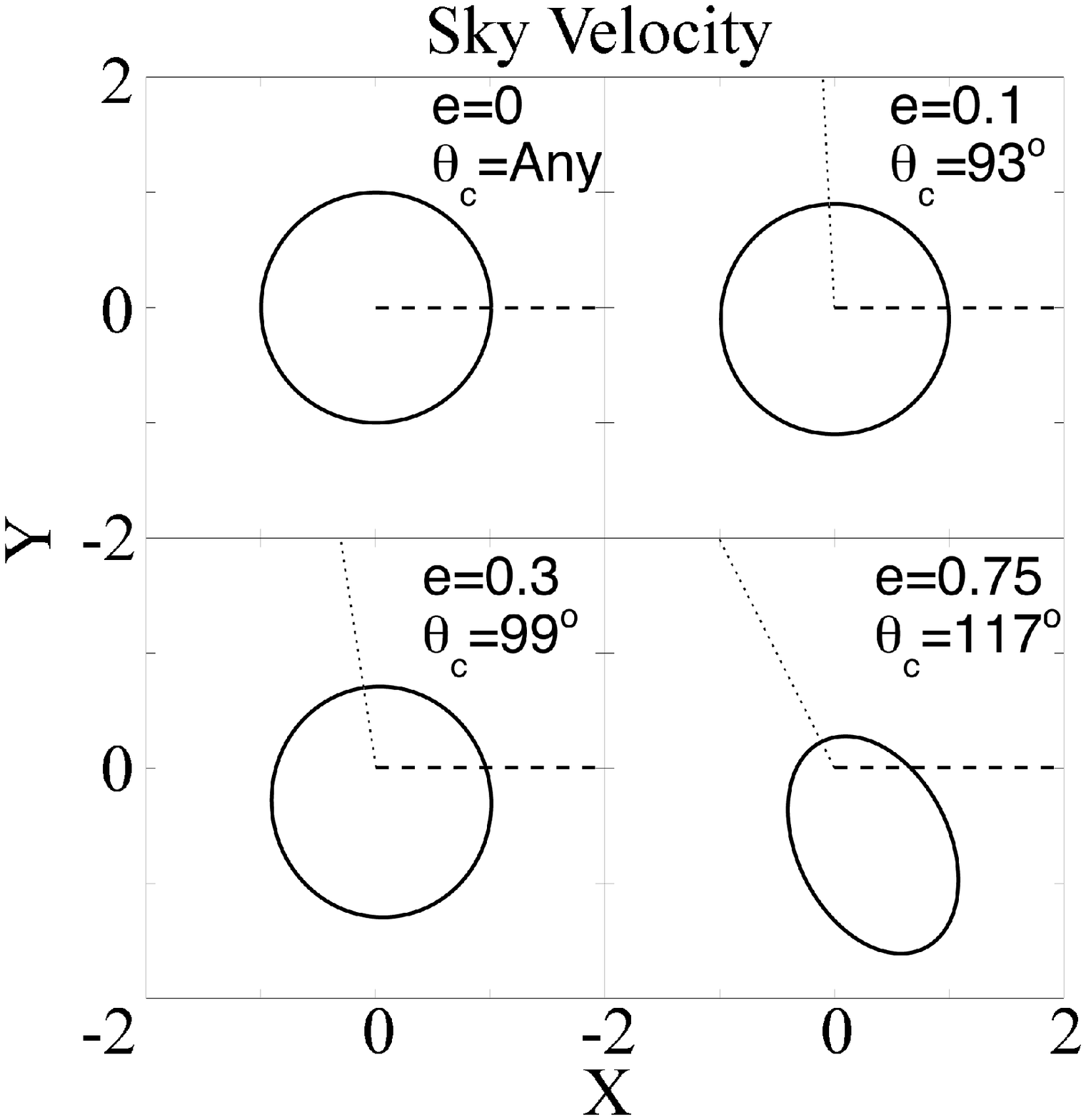}
\end{tabular}
\caption{Locations on an orbit where a velocity is equal to the circular velocity. In all cases, the orbit's semi-major axis is 1, and the observer is assumed to be located at $x=+\infty$, as shown by the dashed line. The longitude of periastron is shown by the dotted line. In each panel the eccentricity is listed, as is one of the values of the true anomaly at which the velocity is circular. In all cases, the orbit is rotated so that that longitude, $\theta_c$, lies on the $+x$-axis. \textit{Left:} The total velocity of the planet is equal to the circular velocity when it crosses the $+x$ axis. In these cases, the observer will see transit durations longer than that from an identical planet on a circular orbit. \textit{Right:} The azimuthal velocity is equal to the circular velocity when it crosses the $+x$ axis. In these cases, an observer will see a transit duration equal to that of an identical planet, but on a circular orbit, \ie $e_{min} = 0$.} 
\label{fig:thetac}
\end{figure*}

The vertical lines in Fig.~\ref{fig:vtheta} show the values of
$\theta$ at which the velocity is equal to the circular velocity. For
the orbital velocity, the longitude where $v = v_c$ is given by
\begin{equation}\label{eq:theta_orb}
\theta_c^{orb} = \pm\cos^{-1}(-e)
\end{equation}
and for the azimuthal velocity, it lies at 
\begin{equation}\label{eq:theta_az}
\theta_c^{sky} = \pm\cos^{-1}\Big(\frac{\sqrt{1-e^2}-1}{e}\Big).
\end{equation}
In Fig.~\ref{fig:thetac}, I show schematics for 4 orbits. In all cases
an observer at $x=+\infty$ views the transit such that either $v =
v_c$ (left) or $v_{sky} = v_c$ (right).

\begin{figure}[h] 
\centering
\begin{minipage}{2.7in}
\resizebox{2.7in}{!}{\includegraphics{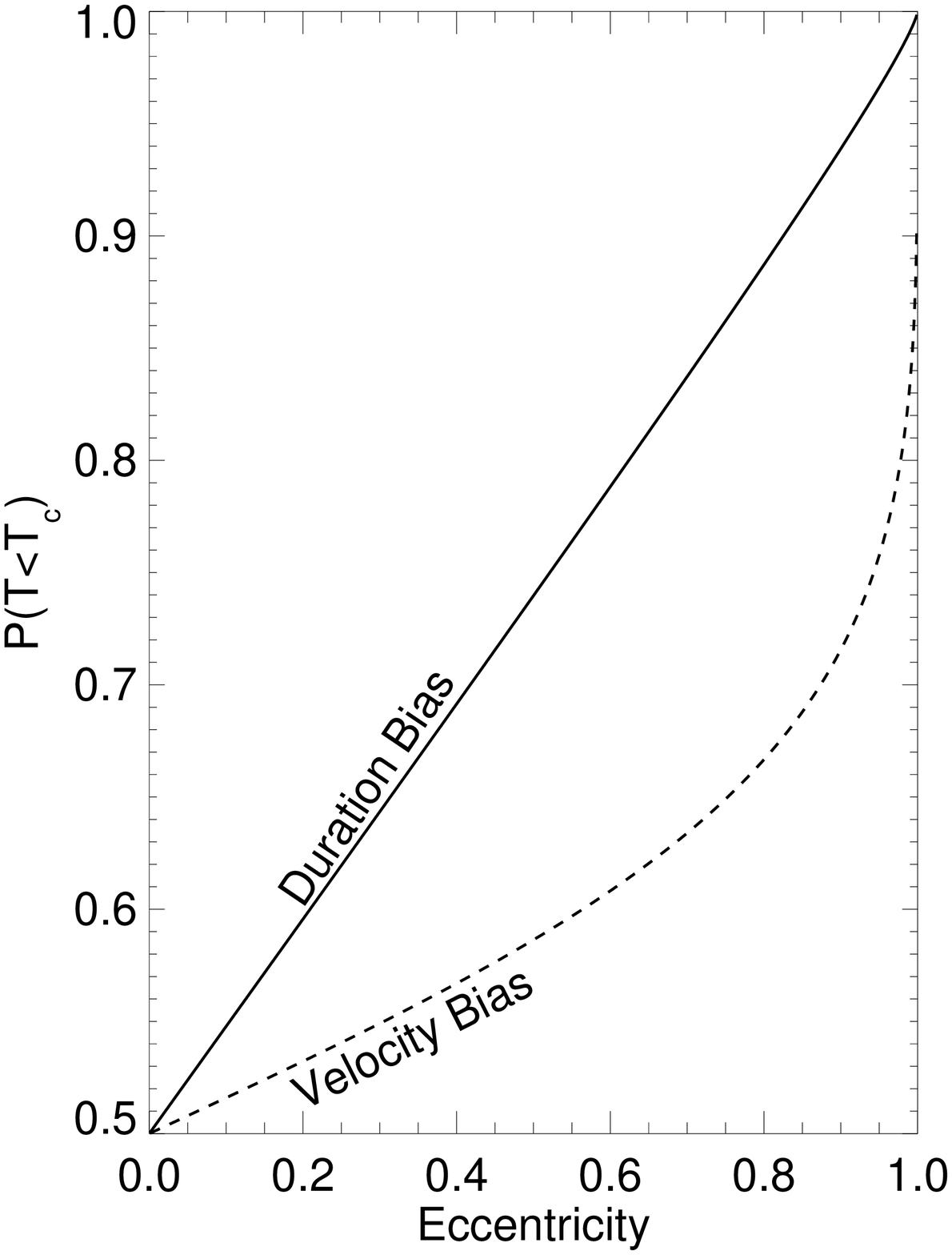}} 
\end{minipage}
\begin{minipage}{2.7in}
\caption{Probability of detecting a transit duration shorter than that of a planet on a circular orbit. The dashed curve is the ``velocity bias'' that arises because more longitudes have $v_{sky} > v_c$. The solid curve is the ``duration bias'' that also takes into account the increased transit probability when the star-planet separation is small. The solid curve is the actual bias, which is not a straight line.} 
\label{fig:prob}
\end{minipage}
\end{figure}

As I am only interested in the azimuthal velocity, for the remainder
of this paper I will assume that $\theta_c = \theta_c^{sky}$ and drop
the superscript. For an eccentric orbit, the planet travels faster
than the circular velocity for $\theta_c/\pi > 0.5$ of the
orbit. There is therefore an observational bias, which I will call the
``velocity bias,'' to observe $\Delta < 1$
\citep{TingleySackett05,Burke08,Plavchan12}. The probability that $T < T_c$ due to this effect is just
\begin{equation}\label{eq:pvel}
p_{vel}(T<T_c) = \frac{\theta_c}{\pi}
\end{equation}
and is shown by the dashed curve in Fig.~\ref{fig:prob}.

The bias toward small $\Delta$ is magnified by the geometrical bias
toward transits occurring at smaller star-planet separations. As shown in J.~W.~Barnes (2007; see
also Borucki \& Summers 1984), the overall transit probability is
\begin{equation}\label{eq:transitprob}
p_{transit} = \frac{1}{4\pi}\frac{2R_*}{a(1-e^2)}\int_0^{2\pi}(1+e\cos\theta)d\theta = \frac{R_*}{a(1-e^2)}
\end{equation}
and the probability to observe the transit when $T < T_c$ is 
\begin{equation}\label{eq:pshort}
p_{short} = \frac{1}{4\pi}\frac{2R_*}{a(1-e^2)}\int_{-\theta_c}^{\theta_c}(1+e\cos\theta)d\theta = \frac{R_*(\theta_c + e\sin\theta_c)}{a\pi(1-e^2)},
\end{equation}
and thus the bias toward observing a transit duration shorter than the circular duration is the ratio of these two equations,
\begin{equation}\label{eq:pduration}
p_{duration} = \frac{\theta_c + e\sin(\theta_c)}{\pi},
\end{equation}
which I will call the ``duration bias.'' This effect is shown by the
solid curve in Fig.~\ref{fig:prob}, and is the actual likelihood to
observe a transit with $T < T_c$. As $e \rightarrow 1$, it becomes
extremely unlikely to observe a long transit. This effect can make
studies that rely on transit durations longer than that predicted for
a central transit ($b=0$) unlikely to find high eccentricity objects
\citep[e.g.][]{Moorhead11,Plavchan12}.

In practice, this bias is not dramatic as most planets are not on very
eccentric orbits. \cite{Burke08} used analytic fits to the
then-current eccentricity distribution as determined from radial
velocity planets, excluding those with orbital periods less than 10
days that may be tidally circularized, to find that the mean value of
$\Delta$ should be 0.88. Recall that \cite{Burke08} used $v$ instead
of $v_{sky}$ in his definition of $\Delta$, thus his mean should be
slightly lower than the actual mean as $v~\ge~v_{sky}$.

\begin{figure*} 
\begin{tabular}{cc}
\includegraphics[width=0.49\textwidth]{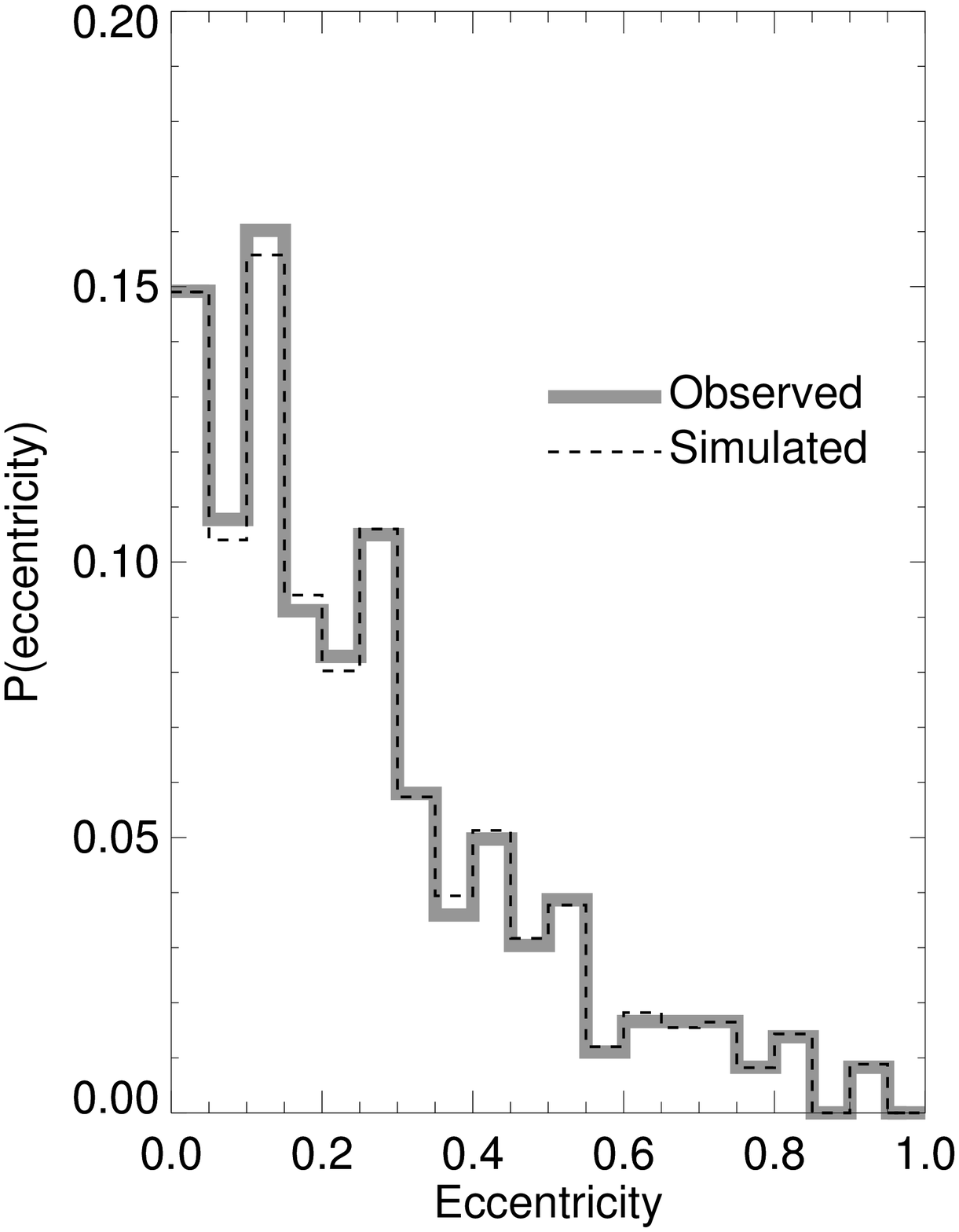} &
\includegraphics[width=0.49\textwidth]{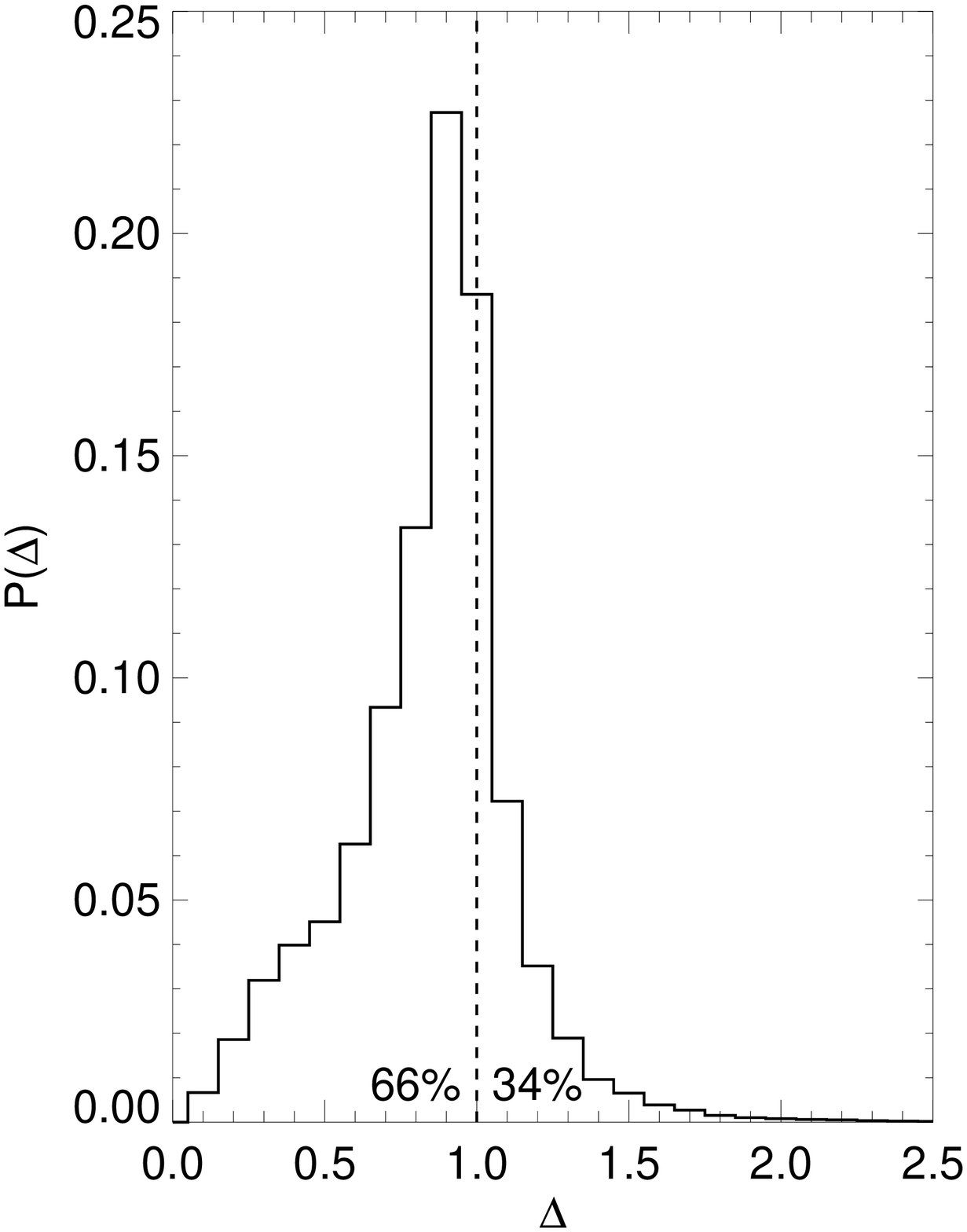}
\end{tabular}
\caption{\textit{Left:} Eccentricity distribution of exoplanets. The solid gray line is the observed distribution (with bin size 0.05) of 362 radial velocity exoplanets with $a > 0.1$~AU from exoplanets.org as of 4 June 2013. The dashed line is the distribution used in this study. \textit{Right:} The transit duration anomaly expected from the simulated data set assuming a random viewing geometry and that the transit duration is set by the azimuthal velocity, Eq.~(\ref{eq:vtheta}).} 
\label{fig:eccdelta}
\end{figure*}

To update the expectations of \cite{Burke08}, I recompute the expected
distribution of $\Delta$ from radial velocity planets. In the left
panel of Fig.~\ref{fig:eccdelta}, I show the current distribution of
$e$ with the thick gray line\footnote{As of 4 June 2013,
http://exoplanets.org}. I exclude those planets with $a>0.1$~AU
leaving 362 planets. To evaluate the expected distribution of
$\Delta$, I created $10^7$ synthetic systems consisting of a star with
a radius between 0.7 and 1.4~R$_\sun$, and a planet whose radius I
ignored. The orbit had $a = 0.05$~AU, an eccentricity distribution
given by the dashed histogram in the left panel of
Fig.~\ref{fig:eccdelta}, and an isotropic distribution of orbits. I
then calculated $\Delta$ for all transiting geometries with durations
larger than 1 hour and show the resulting $\Delta$ distribution in the
right panel of Fig.~\ref{fig:eccdelta}. As noted, 66\% of cases have
$\Delta < 1$, with a mean value of 0.90. If I instead use the circular
velocity to calculate $\Delta$, I find 68\% have $\Delta < 1$, with a
mean of 0.88, reproducing the \cite{Burke08} result. The difference
between using $v_c$ and $v_{sky}$ to calculate the TDA is modest for
the known radial-velocity-detected planets.

\section{Methodology}

In order to determine if the difference in $Q$ values can permit the
identification of $R_{crit}$, I perform Monte Carlo simulations of
both the CPL and CTL models. I then compare the results to
publicly-available \kepler~data to search for the predicted signal.

For my synthetic data, I created 25,000 star-planet configurations
with initial semi-major axes uniformly in the range [0.01,0.15] AU,
planetary radii in the range [0.5,10]~$\rearth$, stellar masses in the
range [0.7,1.4]~$\msun$, a radius in solar radii equal to its mass
in solar masses, and ages in the range [2,8] Gyr. If the planetary
radius is less than $2~\rearth$, then the mass is
$(R/\rearth)^{3.68}\mearth$~\citep{Sotin07}, if larger, then I assume
the density is 1 g/cm$^3$, similar to the planets in the Kepler-11 system \citep{Lissauer13}. The initial eccentricity is drawn from the
currently observed distribution of distant planets ($a > 0.1$~AU), see
Fig.~\ref{fig:eccdelta}. For CPL runs I randomly choose tidal parameters in the ranges $30 \leq Q_r \leq 300$,
$10^6 \leq Q_g \leq 10^7$, and $10^6 \leq Q_* \leq 10^7$. For CTL runs
I used $30 \leq \tau_r \leq 300$~s, $0.003 \leq
\tau_g \leq 0.03$~s, and $0.001 \leq \tau_* \leq 0.01$~s. I then
integrated the system forward for the randomly chosen age and assumed
I observed the system in that final configuration. In order to
calculate $e_{min}$, I choose a random value for $\theta$ that
represents the direction of the observer, and an inclination $i$
chosen uniformly in $\cos i$, with $i$ measured from the plane of the
sky. I then calculate the separation between the star and planet using
\begin{equation}
r = \frac{a(1-e^2)}{1+e\cos\theta}
\label{eq:orbit}
\end{equation}
and determine the impact parameter, $b = r\tan(\pi/2 - i)$. If $b <
R_*$, then the planet transits, and I calculate the transit
duration. As pointed out in \cite{Burke08}, short transit durations
can be missed, and I therefore throw out transit durations that are
less than 1 hour, which is the approximate minimum duration detectable
by \kepler. The vast majority of the rejected transits are too short
due to a large impact parameter, however a few are due to large
eccentricity and alignment of the longitude of pericenter with the
line of sight. Thus, my estimates of the minimum eccentricity
distribution are slightly biased toward lower values. From the
remaining transits, I calculate the TDA and $e_{min}$ using
Eqs.~(\ref{eq:vtheta}--\ref{eq:emin}).

I also compute $e_{min}$ for \kepler~candidates that have all the
requisite parameters presented in \kepler~Planet Candidate Data
Explorer\footnote{http://planetquest.jpl.nasa.gov/kepler}~\citep[see
also][]{Batalha13}.  These data do not contain error bars and
assuredly contain some false positive, but the data set is uniform and
sufficient for this proof of concept. I limit my sample to those with
orbital periods less than 15 days, but will refer to this subsample as
the ``\kepler~sample'' in the upcoming sections.

\section{Results}

\subsection{Tides and the Minimum Eccentricity}

\begin{figure*} 
\begin{tabular}{cc}
\includegraphics[width=0.49\textwidth]{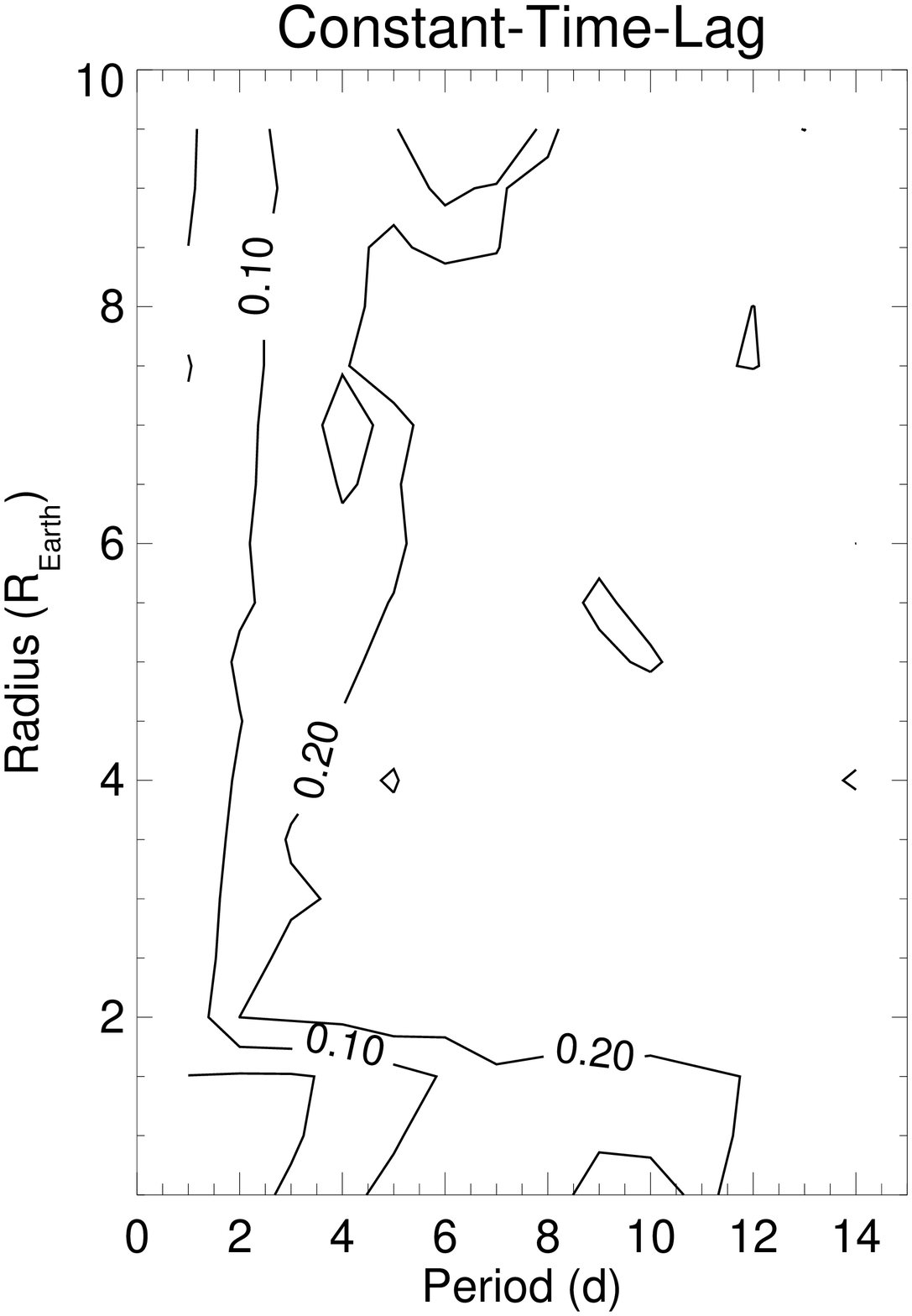} &
\includegraphics[width=0.49\textwidth]{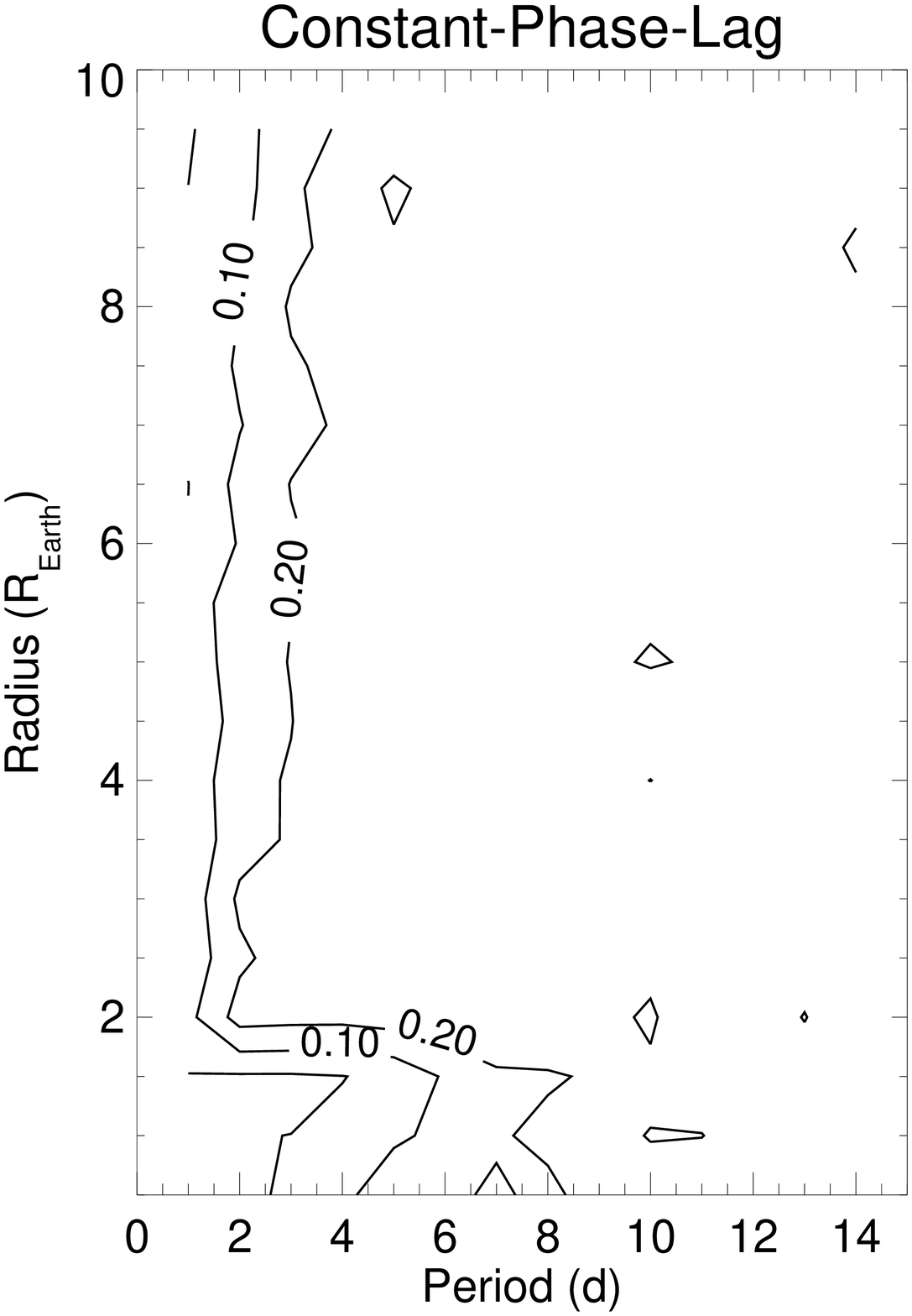}
\end{tabular}
\caption{Average final eccentricity of a suite of $\sim 23,000$
systems of one star and one planet that experienced tidal evolution
for 2--8 Gyr. Planets larger than $2~\rearth$~are gaseous, and those
smaller are rocky. The binsizes are $0.5~\rearth$~ in radius and 0.5 d
in period.  Gaseous planets retain a residual eccentricity if $P >
1.5$ d, while rocky planets require $P > 3$ d. The transition occurs
at $2~\rearth$, which in this case is $R_{crit}$. \textit{Left:}
Orbits evolve according to the constant-time-lag
model. \textit{Right:} Orbits evolve according to the
constant-phase-lag model. }
\label{fig:radper}
\end{figure*}

I begin by considering an intermediate step: In Fig.~\ref{fig:radper},
I show the average final eccentricity of my simulated planets as a
function of planetary radius, $R_p$, and orbital period, $P$.  The
paucity of eccentric orbits at low $R_p$ and $P$ shows the more
effective circularization of rocky bodies.  Furthermore, I can see the
features that correspond directly to three parameters that are
currently very poorly constrained: $R_{crit}$ via the rapid rise in
$<e>$ at the imposed value of $2~\rearth$; $Q_g$ ($\tau_g$) via the
rapid rise in $<e>$ at 1 day above $2~\rearth$; and $Q_r$ ($\tau_r$)
via the rise over 2--10 days and below $2~\rearth$. Thus, despite the
order of magnitude uncertainty I gave to each physical parameter, the
large discrepancy between $Q_r$ ($\tau_r$) and $Q_g$ ($\tau_g$) does
result in an important difference in the expected orbits of close-in
planets of FGK stars. For the CTL model, 1,955 planets merged with the
star during the integrations; 1,118 for CPL
\citep[see][]{Jackson09,Levrard09}.

\begin{figure*} 
\begin{tabular}{cc}
\includegraphics[width=0.49\textwidth]{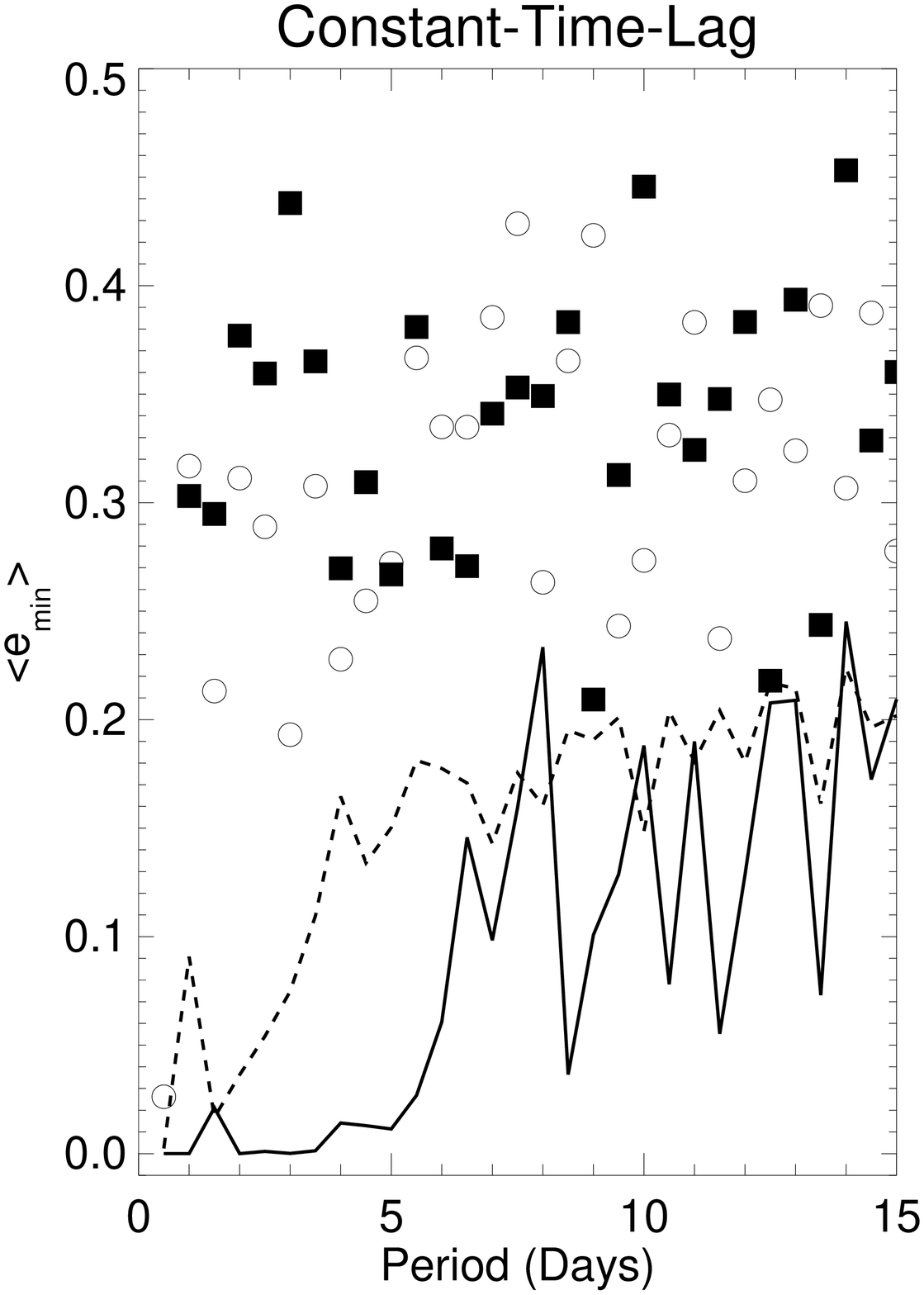}
\includegraphics[width=0.49\textwidth]{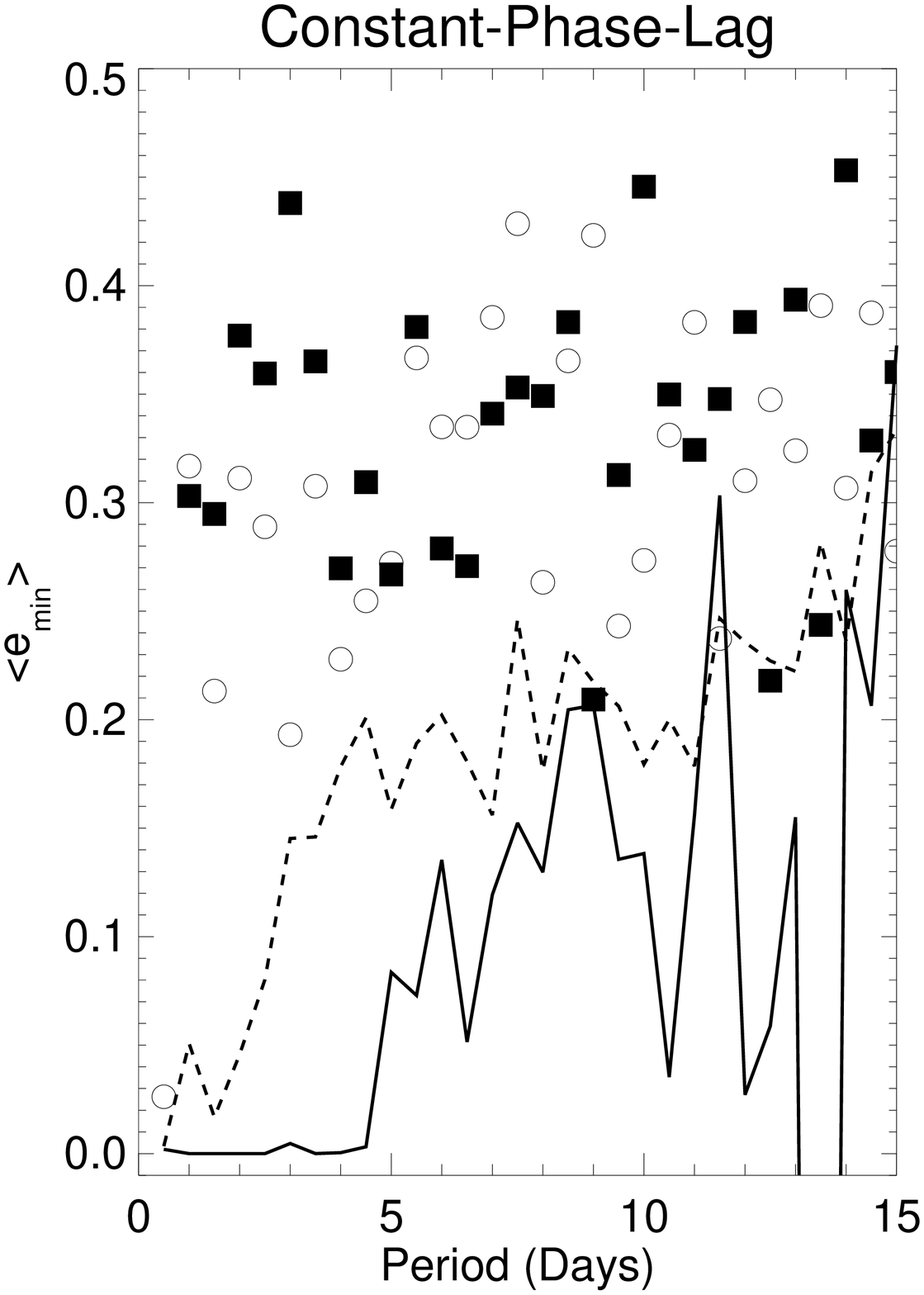}
\end{tabular}
\caption{Average minimum eccentricities for transiting exoplanets
as a function of period and radius.  The solid curve and filled
squares represent planets with radii below the selected critical
radius of $2~\rearth$, while the dashed curve and open circles are
larger planets. The lines are for the simulated data set incorporating
the duration bias, symbols for KOIs. The latter have values of
$<e_{min}>$ of 0.2--0.5 regardless of period, whereas model gaseous
planets have non--zero eccentricities if the period is larger than 1.5
days, and rocky planets at larger than 4. If tidal dissipation is a
function of exoplanet radius, it should be detectable. \textit{Left:}
Results for the CTL model. \textit{Right:} Results for the CPL model.}
\label{fig:emin}
\end{figure*}

Next I calculate the average minimum eccentricity $<e_{min}>$ for
transiting geometries of simulated rocky and gaseous planets in 0.5
day orbital period intervals and plot $<e_{min}>$ as a function of
orbital period for different radii as solid lines in
Fig.~\ref{fig:emin}. For the CTL model, I obtained 2,127
observable transits, and for CPL 2,151, about twice as many
planets as in the same period range as the \kepler~sample. Note that
my synthetic data do not share several properties of the \kepler~data,
such as the planetary radius and period distributions. For both the
CTL (left) and CPL (right) models, the trends are the same.  For $R <
R_{crit}$, $<e_{min}> \sim 0$ up to about a 4--5 day period. However,
for larger radii, circular orbits are only guaranteed for periods less
than about 1.5--2 days. The rocky and gaseous distributions become
about equal at $P = 13$ days, albeit with considerable scatter. At
large orbital periods, the simulated data become sparser as the
transit probability is dropping (producing the apparent oscillations
in $<e_{min}>$), and those that do transit are more likely to be near
pericenter of an eccentric orbit, causing the secular growth in
$<e_{min}>$ with period. Note that similar variations are present in
the \kepler~sample.

Figure \ref{fig:emin} also contains the values of $e_{min}$ provided
by the \kepler~team as squares. Solid squares correspond to $R_p <
R_{crit}$, open to $R_p > R_{crit}$. Nearly all the observed data are
above the predictions, in agreement with the results of
\cite{Moorhead11}. Therefore, it does not appear that there is any
signal of tidal evolution in the \kepler~data, regardless of orbital
period!  This result is in stark contrast to radial velocity data that
show clear signs of circularization at small $P$
\citep[\eg][]{Butler06}. Moreover, the average eccentricity in the
Butler et al.~catalog is $\sim 0.25$, which is lower than the vast
majority of {\it minimum} eccentricities derived from \kepler~transits
in this study. As the radial velocity data are older and have been
reproduced by multiple teams, the \kepler~data are more likely to be
incorrect. In the next section I describe several plausible
explanations for the discrepancy.

\subsection{A Closer Look at the \kepler~Sample}

The lack of evidence of tidal evolution in the KOIs suggests there is
an issue with the interpretation of the light curves. In this section
I examine several features of the \kepler~sample and conclude that the
data suffer from a systematic bias. As described in \cite{Batalha13},
transits are fit to the geometric, limb-darkened transit model of
\cite{MandelAgol02}. This model can determine planetary radius and
impact parameter, as well as other parameters that are not relevant to
the current study. The publicly-available data are long cadence, and
hence the transits are not well-sampled. This sparse sampling is most
likely to affect the impact parameter, as the shape of the transit is
crucial to its estimation. Below I show that the impact parameters do
indeed appear to be suspicious.

\begin{figure*} 
\begin{tabular}{cc}
\includegraphics[width=0.49\textwidth]{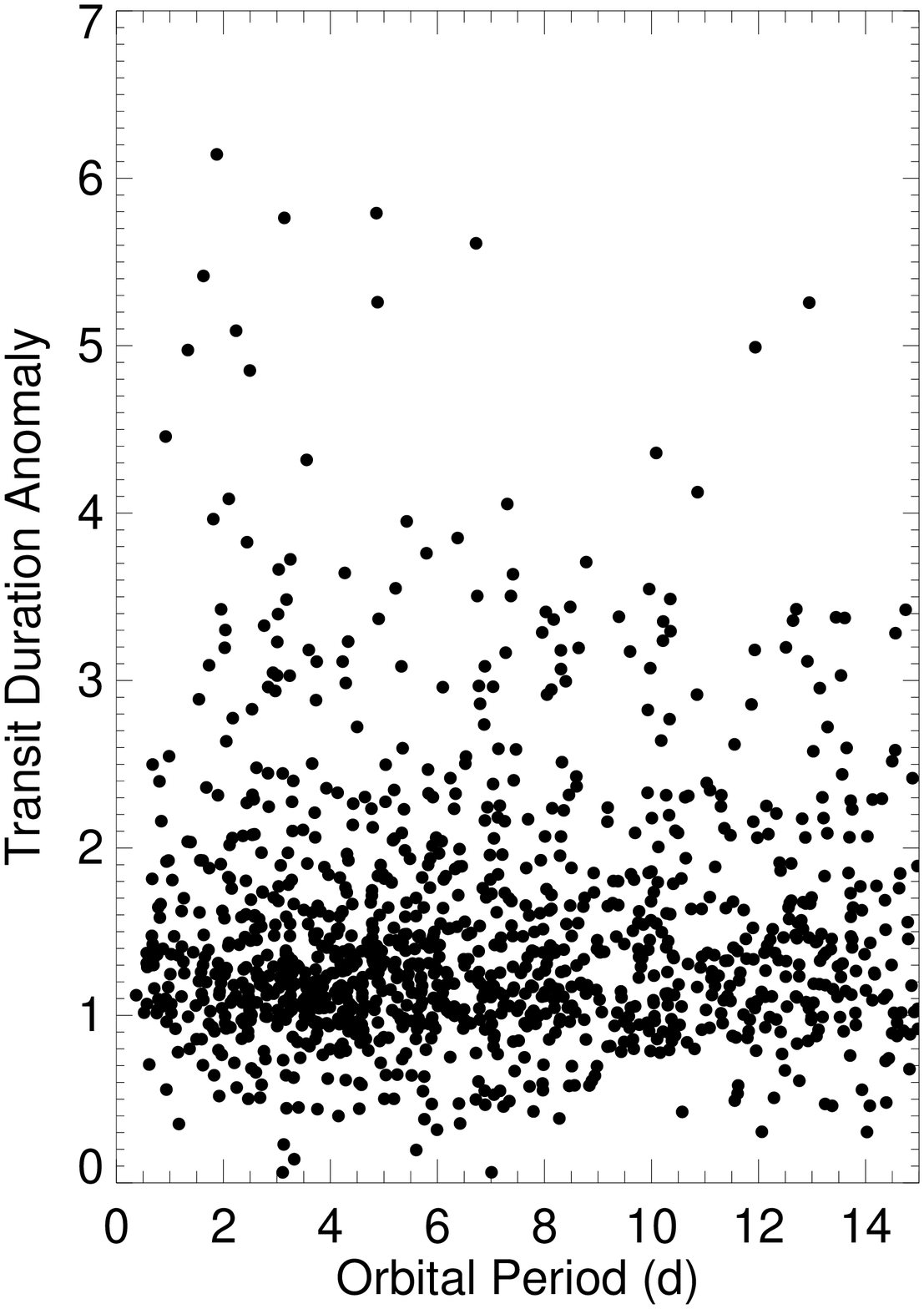}
\includegraphics[width=0.49\textwidth]{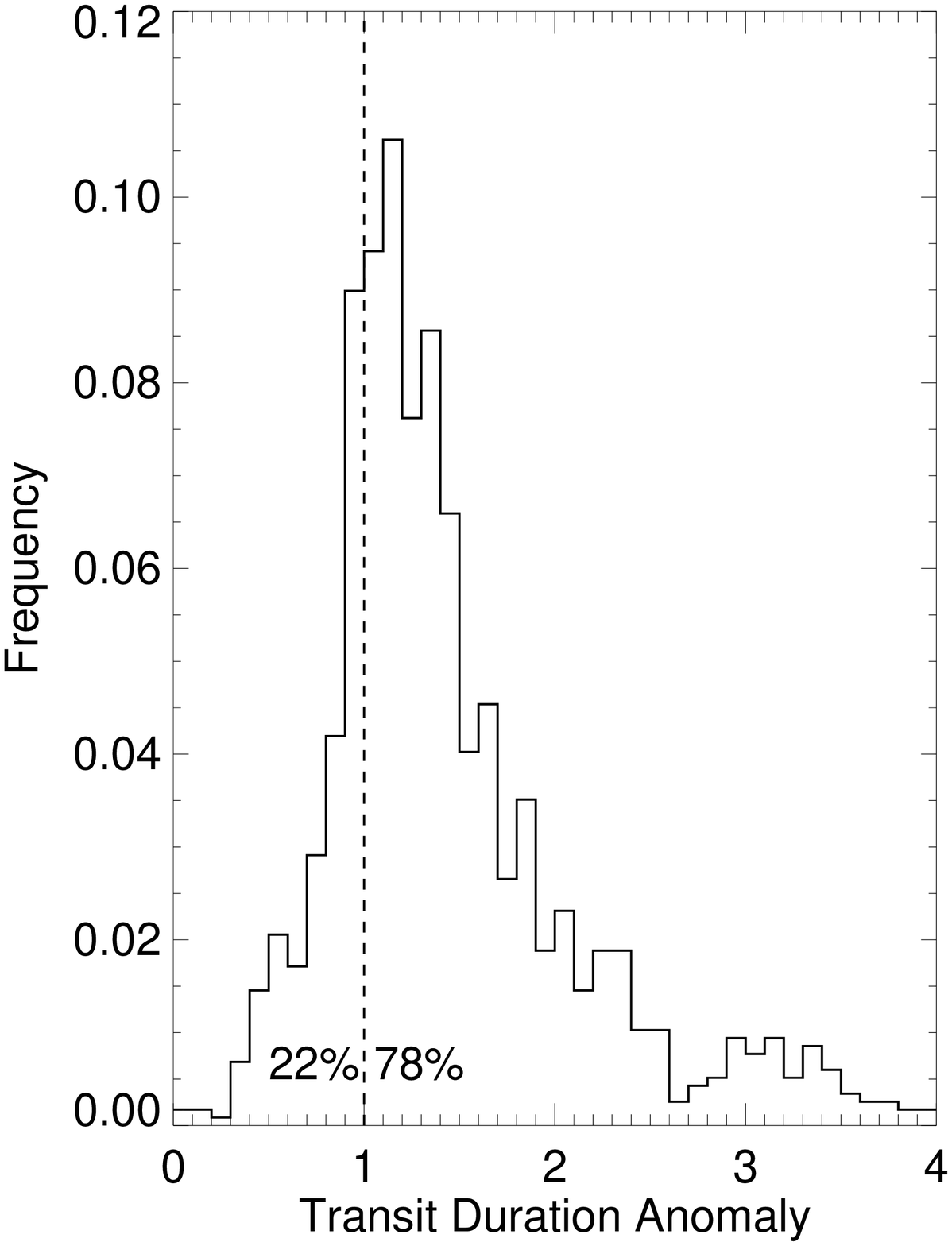}
\end{tabular}
\caption{The transit duration anomalies of the KOIs in the sample (which are short-period). \textit{Left:} TDA as a function of orbital period. \textit{Right:} The histogram of TDA values shows that most values are larger than 1.}
\label{fig:keplertda}
\end{figure*}

A partial list of the \kepler~data used in this study is shown in
Table 1, with the full table available in the supplementary
material\footnote{http://www.astro.washington.edu/users/rory/publications/Barnes14.table1.}. In Fig.~\ref{fig:keplertda} I plot $\Delta$ as a function of
orbital period in the left panel. Although no trends are present, it
does appear that most values of $\Delta$ are greater than 1. In the
right panel I bin the $\Delta$ values to confirm this impression. This
distribution cannot be explained by orbital mechanics and isotropic
orbits which predict that $\Delta$ distributions can only be biased
toward $\Delta < 1$. Instead I find 78\% of KOIs have $\Delta > 1$,
with a mean value of 1.38.

To search for the source of this discrepancy, I considered
relationships among the parameters that permit the calculation of
$e_{min}$. The durations increase monotonically with the orbital
period, albeit with significant scatter, as expected
\citep{Kane12}. Several studies have pointed out systematic errors in
the stellar characterization
\citep{DressingCharbonneau13,Everett13}. In particular,
\cite{Everett13} studied 220 \kepler~host stars and found the vast
majority have larger radii than reported by the \kepler~team, and that
one-quarter are 35\% larger than suggested by the \kepler~team. Since
$\Delta~\propto~R_*^{-1}$, such a revision could significantly lower
$\Delta$ and potentially resolve the discrepancy. Since they ``only''
examined 220 host stars, some of which are not known to host a
close-in planet, I do not include their results here, so that my
analysis is kept to a uniform sample.

\begin{table*}
\begin{center}Table 1. Properties of Short Period KOIs\\
\begin{tabular}{cccccccccc}
\hline\hline
KOI & $a$ (AU) & $P$ (d) & $T$ (hr) & $R_p$ (R$_{Earth}$) & $R_*$
(R$_{Sun}$) & $b$ & $T_c$ (hr) & $\Delta$ & $e_{min}$\\
\hline
1.01 & 0.036 & 2.471 & 1.732 & 14.42 & 1.06 & 0.822 & 1.984 & 0.873 & 0.135\\
2.01 & 0.039 & 2.205 & 3.877 & 22.29 & 2.71 & 0.128 & 5.810 & 0.667 & 0.384\\
3.01 & 0.052 & 4.888 & 2.368 & 4.67 & 0.74 & 0.029 & 2.612 & 0.907 & 0.098\\
4.01 & 0.056 & 3.849 & 2.928 & 11.79 & 2.60 & 0.946 & 2.764 & 1.059 & 0.058\\
5.01 & 0.058 & 4.780 & 2.012 & 5.65 & 1.42 & 0.951 & 1.716 & 1.172 & 0.158\\
5.02 & 0.075 & 7.052 & 3.688 & 0.66 & 1.42 & 0.750 & 3.169 & 1.164 & 0.151\\
7.01 & 0.044 & 3.214 & 4.111 & 3.72 & 1.27 & 0.714 & 2.431 & 1.691 & 0.482\\
10.01 & 0.047 & 3.522 & 3.198 & 15.88 & 1.56 & 0.640 & 3.682 & 0.869 & 0.140\\
17.01 & 0.045 & 3.235 & 3.602 & 11.06 & 1.08 & 0.029 & 3.015 & 1.195 & 0.176\\
18.01 & 0.052 & 3.548 & 4.081 & 17.37 & 2.02 & 0.006 & 5.282 & 0.773 & 0.252\\
20.01 & 0.056 & 4.438 & 4.671 & 17.58 & 1.38 & 0.018 & 4.338 & 1.077 & 0.074\\
...\\
\end{tabular}
\end{center}
\end{table*}

\begin{figure}[h] 
\centering
\begin{minipage}{2.7in}
\resizebox{2.7in}{!}{\includegraphics{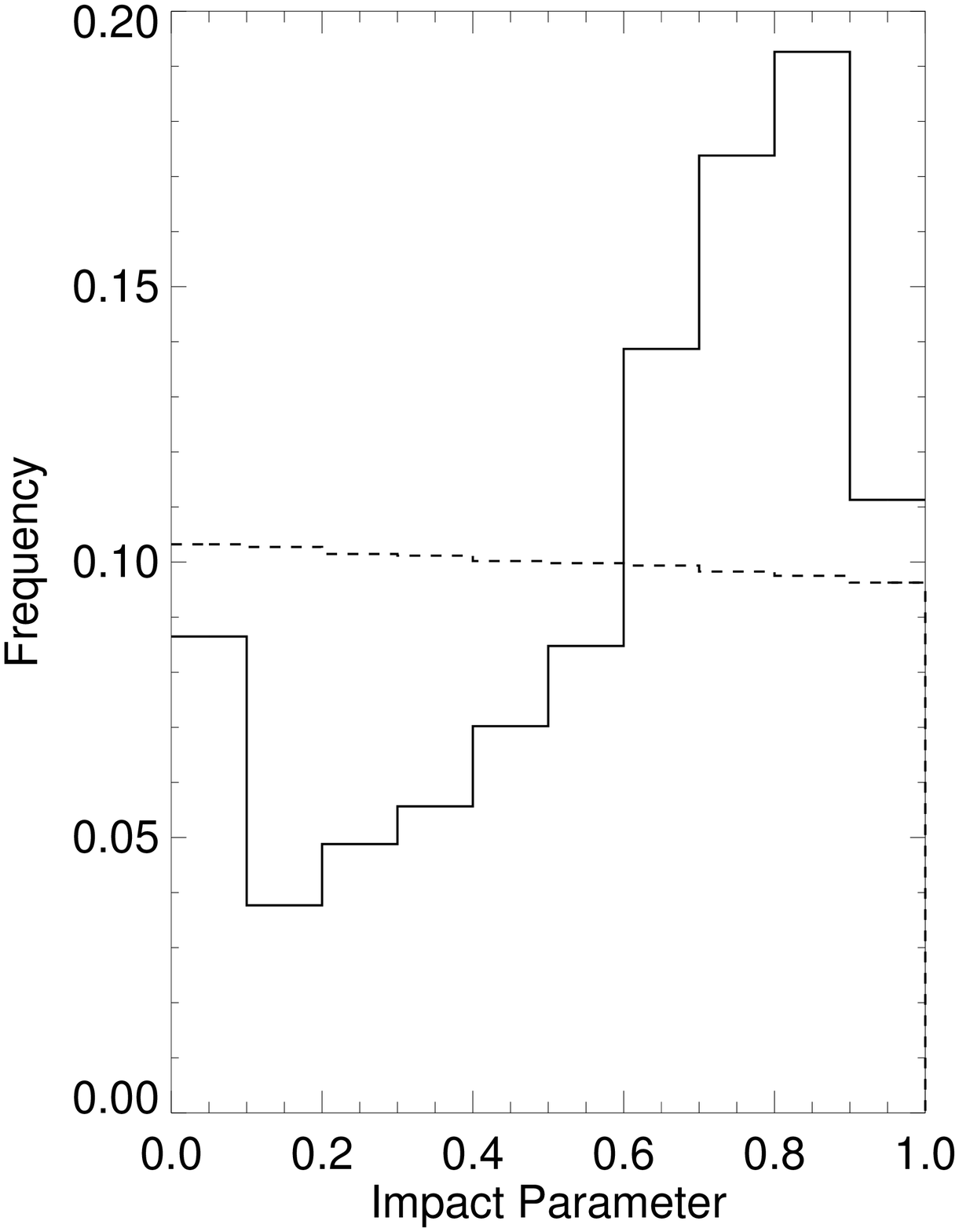}} 
\end{minipage}
\begin{minipage}{2.7in}
\caption{Distribution of impact parameters. The expected distribution of the radial velocity exoplanets with no tidal circularization is shown by the
dashed line. The short-period KOIs tend to have large impact
parameters.}
\label{fig:bhist}
\end{minipage}
\end{figure}

Perhaps the biggest inconsistency in the \kepler~data lies in the
impact parameter, see Fig.~\ref{fig:bhist}. The distribution predicted
by the radial velocity exoplanets beyond the reach of tides is shown
with the dashed histogram and is taken from the same sample that
produced the right panel of Fig.~\ref{fig:eccdelta}. It is
approximately flat, with the slight rise toward small values due to
isotropically distributed orbits favoring edge-on geometries. Instead,
the \kepler~sample, shown by the solid line, rises sharply to large
values of $b$. This distribution hints that a systematic error may be
present in the \kepler~analysis, which manifests itself in my analysis
into large values of $\Delta$ and $e_{min}$. I conclude that the
currently available \kepler~data produce unreliable values of $b$ and
hence $e_{min}$.

\section{Discussion}

My simulated data show that $R_{crit}$ is identifiable in transit data
due to the difference in the tidal $Q$s of gaseous and rocky bodies,
at least for my idealized model of exoplanetary properties. My
analysis is somewhat circular as I split the data in
Fig.~\ref{fig:emin} on the selected value of $R_{crit}$. In reality,
its value is unknown and must be searched for. However, given the
large and approximately equal values of $<e_{min}>$ in the
\kepler~data, I did not perform that search. More accurate data are
needed, and may be available as KOI parameters are refined. Resolution
of transit ingress and egress may be possible with short cadence data
(which are unpublished but assuredly a small fraction of the total
number of KOIs), or by folding the hundreds of transits together
\cite[e.g.,][]{Jackson13}, potentially rectifying the discordance
between the \kepler~and radial velocity data.

In order to accurately determine the minimum eccentricities, one needs
reliable information for both $R_*$ and $b$, but it is not yet
available.  Although subsets of more reliable data are available for
the former \citep[\eg][]{Everett13}, the transit fits are still
plagued by inaccurate calculations of the impact
parameter. Determination of the impact parameters in short cadence
data or by folding would require a new and comprehensive analysis of
those light-curves and is beyond the scope of this study. After those
data have been properly analyzed, the technique described in this
study should be re-applied in order to determine $R_{crit}$, $Q_g$,
and $Q_r$.

Aside from systematic errors in the analysis of the light curves,
physical effects can also impact the value of $<e_{min}>$. First, I
note that additional companions can pump eccentricity through mutual
gravitational interactions, even if tidal damping is ongoing
\citep{MardlingLin02,Bolmont13}.  Therefore one must be cautious when
interpreting Fig.~\ref{fig:emin}, as additional companions, both seen
and unseen, can maintain non--zero eccentricities.  However,
\cite{Bolmont13} showed that planet--planet interactions cannot
maintain the eccentricity of the hot super-Earth 55 Cnc e above 0.1.
That system is particularly relevant as there are many close--in
planets orbiting a typical G dwarf.  Therefore, I conclude that
eccentricity pumping can be significant, but cannot explain the
discrepancy between the observed and simulated systems shown in
Fig.~\ref{fig:emin}. This analysis should be revisited when all
the \kepler~data become available and all issues with host star
characterization have been resolved.

Another possibility is that stellar winds and activity can strip an
atmosphere, reducing the mass and radius, and potentially changing the
planet from a mini--Neptune to a super--Earth
\citep{Jackson10,Valencia10,Leitzinger11,Poppenhaeger12}.  Recently,
\cite{OwenWu13} argued that the \kepler~sample is consistent with
hydrodynamic mass loss, and that some low--mass planets could have
formed with substantially more mass.  Mass loss should increase the
time to circularize the orbit, assuming the radius doesn't become very
large, which is unlikely after about 100 Myr \citep{Lopez12}.
Therefore, mass loss could stall circularization for mini--Neptunes,
but not for super--Earths.  Although few radial velocity measurements
exist for planets with radii less than $\sim 1.5~\rearth$, they have
densities consistent with silicate compositions \citep{Batalha11}.
Thus, mass loss seems unlikely to explain the differences seen for the
smallest candidates in the \kepler~field.

Radial inflation by irradiation \citep[\eg][]{Lopez12} or tidal
heating \citep{Bodenheimer01,Jackson08_heat,IbguiBurrows09} also work
to decrease $e$ since the evolution scales as $R_p^5$. Hence, bloated
planets should be found on circular orbits, but no such trend is
observed in the \kepler~candidates.

In this study I used two qualitatively different equilibrium
tidal models and standard assumptions for dissipation. However
different tidal models have been
proposed \citep[e.g.][]{OgilvieLin04,Henning09,Socrates12,MakarovEfroimsky13}
and could be applied to this problem. The trend I predict
here holds unless the dissipation rates in gas giants and rocky
planets are within 1--2 orders of magnitude of each other, rather
4--6. Recently, \cite{StorchLai13} have proposed just such a model in
which all tidal dissipation in gas giants occurs in a rocky
core. Should all planets show the same trend in $e_{min}$, then that
would be evidence in support of their model. Hence, even if the
expectations laid out above prove to be incorrect, some tidal models
could be rejected by the methodology used in this study.

I have focused on transiting planets, but an analogous study could be
applied to radial velocity data. Those data may be more amenable to
such a study as orbital eccentricity is a direct observable. The
problem lies in the low reflex velocities induced by the small planets
as well as the ambiguity in mass due to the mass-inclination
degeneracy. Nonetheless, with enough objects and an accounting for the
expected isotropy of orbits, it may be possible to determine tidal
dissipation as a function of mass in radial velocity data.

The \kepler~spacecraft was designed to discover a potentially
habitable planet orbiting a solar-like star. Such a planet ($m_p \lsim
10~\mearth$; $r_p \lsim 2~\rearth$; $P \approx 1$~year) would have an
undetectable radial velocity signature, preventing a direct mass
measurement. Thus, confirmation of that planet's rocky nature is
daunting. However, \kepler~data may also hold the key to a convincing
solution to the problem, as shown in this study. The distribution of
TDAs in conjunction with tidal theory suggests the value of $R_{crit}$
may be calculated from the close-in planets. Although these planets
are not habitable, they may provide crucial information to assess the
habitability of Earth-like planets that transit Sun-like stars.

\section{Conclusions}

I have shown that the expected difference in tidal dissipation between
gaseous and terrestrial exoplanets should lead to tidal
circularization at different orbital distances. I have also shown how
transit data, namely the transit duration anomaly, can be used to
determine the critical radius between gaseous and terrestrial
planets. Moreover, an analysis of $e_{min}$ can also constrain the
tidal dissipation in exoplanets, an understanding of which is sorely
needed. Using standard values for tidal parameters and the critical
radius, I find that a large ensemble of transit data should identify
the critical radius between rocky and gaseous exoplanets. My analysis
of available \kepler~data reveals that the theoretical expectations
are not met. However, this discrepancy cannot be used to refute the
hypothesis because the values of $<e_{min}>$ are inconsistent with
radial velocity detected exoplanets, particularly where tidal damping
has been observed.

I have also reviewed the derivation of the TDA, as well as the known
biases toward small values. Previous studies have advocated different
choices for the velocity of the transiting planet as a function of
true anomaly. The transit duration is actually determined by the
azimuthal velocity, which is the velocity in the sky plane
(Eq.~[\ref{eq:vtheta}]). While it is not clear in all previous studies
which velocity was used, those that used the orbital velocity will
obtain slightly smaller expected values of $\Delta$. I find that the
current distribution of exoplanet eccentricities predicts that 66\% of
transit durations should be less than $T_c$, with a mean and mode of
0.9. In contrast, 78\% of short-period KOIs have durations {\it greater}
than $T_c$ with a mean of 1.38. This distribution is inconsistent with
celestial mechanics and the expectation isotropic orbits, regardless
of tidal damping. 

As the \kepler~data are refined, or as new data, e.g.~from the
\textit{TESS} mission arrive, this hypothesis should be revisited. The
value of $R_{crit}$ is crucial for the interpretation of the
habitability of transiting Earth-sized planets orbiting Sun-sized
stars. Moreover, planets in the habitable zones of M dwarfs are
susceptible to tidal effects
\citep{Dole64,Kasting93,Jackson08_hab,Heller11,Barnes13}, so a
determination of $Q$ for terrestrial exoplanets is also crucial to
assessing habitability of planets such as those orbiting Gl 581
\citep{Udry07,Mayor09,Vogt10} and Gl 667C
\citep{AngladaEscude12,Bonfils13,AngladaEscude13}. As missions like
\kepler~and \textit{TESS} have been designed to find potentially
habitable worlds, the determination of $R_{crit}$ through their data
alone would be an important step forward in determining the occurrence
rate of terrestrial planets in the HZ.

\vspace{1cm}
\noindent\large Acknowledgments\\
\normalsize\noindent I thank Andrew Becker, Eric Agol, Leslie Hebb, Jason Barnes, Brian Jackson and Ren{\' e} Heller for helpful discussions. This work was supported by NSF grant AST-110882 and the NAI's Virtual Planetary Laboratory lead team. I also thank an anonymous referee and Nader Haghighipour for reviews that greatly improved the clarity and accuracy of this manuscript.

\bibliography{CriticalRadius}

\end{document}